\documentclass[aps,prd,reprint,amsmath,amssymb,superscriptaddress,nofootinbib,floatfix,longbibliography]{revtex4-1}
\usepackage[utf8]{inputenc}
\usepackage{graphicx}
\usepackage{dcolumn}
\usepackage{bm}
\usepackage{hyperref}
\usepackage{siunitx}
\usepackage{mathtools}
\usepackage{aas_macros}
\usepackage{enumerate}

\usepackage{natbib}

\usepackage{xcolor}

\newcommand{\gitDate}{2018-08-07 17:19:26 +0200}
\newcommand{\gitID}{347c058}
\newcommand{\gitStatus}{CLEAN}

\newcommand{\FF}{\mathcal{F}}
\newcommand{\DD}{\mathrm{ \, d}}

\renewcommand{\Re}{\operatorname{Re}}

\newcommand{\stat}{d}
\newcommand{\avgFstat}{\overline\FF}
\newcommand{\sumFstat}{\hat\FF}
\newcommand{\cdf}{\operatorname{cdf}}
\newcommand{\pdf}{\operatorname{pdf}}
\newcommand{\median}{\operatorname{median}}
\newcommand{\depth}{\mathcal D}
\newcommand{\Tseg}{T_{\mathrm{seg}}}
\newcommand{\TSFT}{T_{\mathrm{SFT}}}
\newcommand{\NSFT}{N_{\mathrm{SFT}}}
\newcommand{\eff}{_{\mathrm{eff}}}
\newcommand{\loudest}{^{*}}

\newcommand{\rhoOpt}{\rho_0}
\newcommand{\rhoEff}{\rho\eff}
\newcommand{\rhoEffPerSeg}{\rho_{\mathrm{eff},\ell}}
\newcommand{\Sn}{S_{\mathrm{n}}}
\newcommand{\scalar}[2]{\left(#1,#2\right)}
\newcommand{\Tdata}{T_{\mathrm{data}}}
\newcommand{\uvec}[1]{\hat{#1}}
\newcommand{\vSky}{\uvec{n}}
\newcommand{\al}{\alpha}
\newcommand{\avg}[1]{\left\langle#1\right\rangle}
\newcommand{\pDET}{p_{\mathrm{det}}}
\newcommand{\pFA}{p_{\mathrm{fa}}}
\newcommand{\pFD}{p_{\mathrm{fd}}}
\newcommand{\pTH}{p\thr}
\newcommand{\Fth}{\FF\thr}
\newcommand{\prob}[2]{P(#1\mid#2)}
\newcommand{\prior}[1]{P(#1)}
\newcommand{\Nseg}{N_{\mathrm{seg}}}
\newcommand{\mis}{\mu}
\newcommand{\CO}[1]{\tilde{#1}}
\newcommand{\SC}[1]{\hat{#1}}
\newcommand{\misCO}{\CO{\mis}}
\newcommand{\misSC}{{\mis}}
\newcommand{\FCO}{\CO{\FF}}
\newcommand{\FSC}{\SC{\FF}}
\newcommand{\Fmax}{\FF\loudest}
\newcommand{\thr}{_{\mathrm{th}}}
\newcommand{\stage}[1]{^{(#1)}}
\newcommand{\nStages}{n}
\newcommand{\total}{^{(\mathrm{tot})}}
\newcommand{\nc}{n_{\mathrm{c}}}
\newcommand{\ncTH}{n_{\mathrm{c,th}}}
\newcommand{\parSpace}{\mathbb{P}}
\newcommand{\templBank}{\mathbb{T}}
\newcommand{\Ntempl}{\mathcal{N}}
\newcommand{\single}{^1}
\newcommand{\signal}{_\mathrm{s}}
\newcommand{\Gauss}{\mathrm{Gauss}}
\newcommand{\Amp}{\mathcal{A}}
\newcommand{\M}{\mathcal{M}}
\newcommand{\plus}{_+}
\newcommand{\cross}{_\times}
\newcommand{\eps}{\varepsilon}
\newcommand{\Hz}{\text{Hz}}
\newcommand{\perRootHz}{\Hz^{-1/2}}
\newcommand{\meas}{_{\mathrm{meas}}}
\newcommand{\est}{_{\mathrm{est}}}
\newcommand{\best}{_{\mathrm{best}}}
\newcommand{\worst}{_{\mathrm{worst}}}
\newcommand{\Ord}[1]{\mathcal{O}\left(#1\right)}
\newcommand{\Fstat}{$\FF$}
\newcommand{\Gstat}{$\mathcal{G}$}
\newcommand{\Coinc}{+Coinc}
\newcommand{\StackSlideF}{StackSlide-$\FF$}
\newcommand{\HoughF}{Hough-$\FF$}
\newcommand{\TDFstat}{\Fstat$_{\mathrm{TD}}$\Coinc}
\newcommand{\GCT}{\StackSlideF}
\newcommand{\BayesPE}{BayesPE}
\newcommand{\FollowUp}{FUP}
\newcommand{\PF}{^{\mathrm{PF}}}

\newcommand{\depthMED}{\depth^{\mathrm{med}}}
\newcommand{\sigmaMAD}{\widehat{\sigma}}

\newcommand{\AllSky}{AllSky}

\newcommand{\search}[3]{\texttt{#1-#2-#3}}
\newcommand{\method}[1]{\texttt{#1}}

\newcommand{\ninety}{\textit}

\newcommand{\dcc}{\href{https://dcc.ligo.org/P1800198-v3}{LIGO-P1800198-v3}}

\begin{document}
\title{Fast and Accurate Sensitivity Estimation for Continuous-Gravitational-Wave Searches }

\date{\gitDate, commit \gitID-\gitStatus, \dcc}
\author{Christoph Dreissigacker}
\email{christoph.dreissigacker@aei.mpg.de}
\author{Reinhard Prix}
\affiliation{Max-Planck-Institut für Gravitationsphysik (Albert-Einstein-Institut), D-30167 Hannover, Germany}
\affiliation{Institut für Gravitationsphysik, Leibniz Universität Hannover, D-30167 Hannover, Germany}
\author{Karl Wette}
\affiliation{ARC Centre of Excellence for Gravitational Wave Discovery (OzGrav) and Centre for Gravitational Physics, Research School of Physics and Engineering, The Australian National University, Acton ACT 2601, Australia}
\affiliation{Max-Planck-Institut für Gravitationsphysik (Albert-Einstein-Institut), D-30167 Hannover, Germany}
\affiliation{Institut für Gravitationsphysik, Leibniz Universität Hannover, D-30167 Hannover, Germany}

\begin{abstract}
  This paper presents an efficient numerical sensitivity-estimation method and implementation for
  continuous-gravitational-wave searches, extending and generalizing an earlier analytic approach by
  Wette~\cite{wette2012estimating}.
  This estimation framework applies to a broad class of $\FF$-statistic-based search methods, namely
  (i) semi-coherent StackSlide $\FF$-statistic (single-stage and hierarchical multi-stage),
  (ii) Hough number count on $\FF$-statistics, as well as
  (iii) Bayesian upper limits on (coherent or semi-coherent) $\FF$-statistic search results.
  We test this estimate against results from Monte-Carlo simulations assuming Gaussian noise.
  We find the agreement to be within a few $\%$ at high (i.e.\ low false-alarm) detection thresholds, with increasing
  deviations at decreasing (i.e.\ higher false-alarm) detection thresholds, which can be understood in terms of the
  approximations used in the estimate.
  We also provide an extensive summary of sensitivity depths achieved in past continuous-gravitational-wave searches (derived from the
  published upper limits). For the $\FF$-statistic-based searches where our sensitivity estimate is applicable, we
  find an average relative deviation to the published upper limits of less than $10\%$, which in most cases includes
  systematic uncertainty about the noise-floor estimate used in the published upper limits.
\end{abstract}

\maketitle

\section{\label{sec:intro} Introduction}
The recent detections of gravitational waves from merging binary-black-hole and double neutron-star systems
\cite{GW150914,2016arXiv160604856T,GW170817} have opened a whole new observational window for astronomy, allowing for
new tests of general relativity \cite{2016arXiv160203841T}, new constraints on neutron star physics
\cite{2018arXiv180511581T} and new measurements of the Hubble constant \cite{2017Natur.551...85A}, to mention just a few
highlights.

Continuous gravitational waves (CWs) from spinning non-axisymmetric neutron stars represent a different class of
potentially-observable signals \cite{prix2009gravitational,2015PASA...32...34L}, which have yet to be detected
\cite{2017MPLA...3230035R}. These signals are expected to be long-lasting (at least several days to years)
and quasi monochromatic, with slowly changing (intrinsic) frequency.
The signal amplitude depends on the rich (and largely not yet well-understood) internal physics
of neutron stars \cite{2013PhRvD..88d4004J}, as well as their population characteristics
\cite{camenzind2007compact,2008PhRvD..78d4031K}.
A detection (and even non-detection) of CWs could therefore help us better understand these
fascinating astrophysical objects, and may allow for new tests of general relativity
\cite{2017PhRvD..96d2001I,2018PhRvL.120c1104A}.

\subsection*{Overview of search categories}
\label{sec:categ-cw-search}

We can categorize CW searches in two different ways: either based on the \emph{search method}, or on the type of
\emph{explored parameter space}.

The \emph{search methods} fall into two broad categories: \emph{coherent} and \emph{semi-coherent} (sometimes also
referred to as incoherent). Roughly speaking, a coherent search is based on signal templates with coherent phase
evolution over the whole observation time, while semi-coherent searches typically break the data into shorter coherent
segments and combine the resulting statistics from these segments incoherently (i.e.\ without requiring a consistent
phase evolution across segments). However, there are many different approaches and variations, which are beyond the scope
of this paper, see e.g.\ \cite{2017MPLA...3230035R} for a more detailed overview.
Here we will exclusively focus on coherent and semi-coherent methods based on the $\FF$-statistic, which will be
introduced in Sec.~\ref{sec:searches}.

Coherent search methods are the more sensitive in principle, but in practice they usually suffer from severe computing-cost
limitations: for finite search parameter spaces the required number of signal templates typically grows as a steep power-law
of the observation time, making such searches infeasible except when the search region is sufficiently small.
For larger signal parameter spaces the observation time needs to be kept short enough for the search to be
computationally feasible, which limits the attainable coherent sensitivity. This is where semi-coherent searches tend to
yield substantially better sensitivity at fixed computing cost (e.g.\ see \cite{2000PhRvD..61h2001B,prix2012optimal}).

Based on the \emph{explored parameter space}, we distinguish the following search categories (referencing a recent
example for each case):

\begin{enumerate}[(i)]

\item \emph{Targeted searches} for known pulsars \cite{ATNF} assume a perfect fixed relationship between the observed neutron-star spin
frequency and the CW emission frequency. Therefore one only needs to search a single point in parameter space for each
pulsar, allowing for optimal fully-coherent searches \cite{O1KnownPulsar}.

\item \emph{Narrow-band searches} for known pulsars assume a small uncertainty in the relationship between CW
frequency and the measured pulsar spin rates. This finite search parameter space requires a template bank with
(typically) many millions of templates, still allowing for optimal fully-coherent search methods to be used \cite{O1Narrowband}.

\item \emph{Directed (isolated) searches} aim at isolated neutron stars with known sky-position and unknown spin
frequency. The search parameter space covers the unknown frequency and spindowns of the neutron star
signal within an astrophysically-motivated range \cite{S6NineYoung,S6CasA}.

\item \emph{(Directed) binary searches} aim at binary systems with known sky-position and parameter-space uncertainties
in the frequency and binary-orbital parameters. Typically these sources would be in low-mass X-ray binaries, with the
most prominent example being Scorpius X-1 (Sco~X-1)) \cite{O1ScoX1CrossCorr, O1ScoX1}.

\item \emph{All-sky (isolated) searches} search the whole sky over a large frequency (and spindown) band for unknown
isolated neutron stars \cite{O1AS,O1ASE@H}.

\item \emph{All-sky binary searches} are the most extreme case, covering the whole sky for unknown neutron stars in
binary systems \cite{goetz2011all,S6Binary}.

\end{enumerate}

\subsection*{Sensitivity estimation}
\label{sec:sens-estim}

In this work we use the term \emph{sensitivity} to refer to the upper limit on \emph{signal ampltitude} $h_0$ (or
equivalently \emph{sensitivity depth} $\depth\equiv\sqrt{\Sn}/h_0$, see Sec. \ref{sec:sensitivity-depth}). This can be either the frequentist upper limit for
a given detection probability at a fixed false-alarm level (p-value), or the Bayesian upper limit at a given credible
level for the given data.

Sensitivity therefore only captures one aspect of a search, namely how ``deep'' into the noise-floor it can
detect signals, without accounting for how ``wide'' a region in parameter space is covered, how much
prior weight this region contains, or how robust the search is to deviations from the signal model.
Comparing sensitivity depth therefore only makes sense for searches over very similar parameter spaces.
A more complete measure characterizing searches would be their respective detection probability, which folds in
sensitivity depth, breadth in parameter space, as well as the prior weight contained in that space \cite{2016PhRvD..93f4011M,2018PhRvD..97b4051M}.

However, it is often useful to be able to reliably and cheaply estimate the sensitivity of a search setup without
needing expensive Monte-Carlo simulations:
\begin{itemize}
\item In order to determine optimal search parameters for a semi-coherent search (i.e.\ the number and semi-coherent
segments and template-bank mismatch parameters), it is important to be able to quickly asses the projected sensitivity
for any given search-parameter combination (e.g.\ see \cite{prix2012optimal,2015arXiv150200914L,2016PhRvD..93f4011M,2018PhRvD..97b4051M}).

\item For setting upper limits for a given search, one typically has to repeatedly add software-generated CW
signals to the data and perform a search, in order to measure how often these signals are recovered above a given
threshold. By iterating this procedure one tries to find the weakest signal amplitude that can be recovered at the
desired detection probability (or ``confidence''). This can be very computationally expensive, and a quick and
reasonably-reliable estimate for the expected upper-limit amplitude can therefore substantially cut down on the cost of
this iterative process, which can also improve the accuracy of the upper limit.

\item The estimate can also serve as a sanity check for determining upper limits\footnote{In fact, in the course of this
    work we have identified a bug in the upper-limit script of a published result, while trying to understand the
    discrepancy between the estimate and the published value, see Sec.~\ref{ssec: CompDirected}.}.
\end{itemize}
A number of theoretical sensitivity estimates have been developed over the past decades.
One of the first estimates was obtained for a coherent $\FF$-statistic search \cite{S1J1939}, yielding
\begin{equation}
  \label{eq:40}
  h_0 = 11.4 \sqrt{\frac{\Sn}{\Tdata}}\,,
\end{equation}
for a $90\%$ confidence upper limit at $1\%$ false-alarm (per template). $\Sn$ denotes the (single-sided) noise power
spectral density, and $\Tdata$ is the total amount of data.
This was later generalized to the semi-coherent Hough \cite{krishnan2004hough} and StackSlide method
\cite{mendell05:_stackslide,S4AS}, yielding an expression of the form
\begin{equation}
  \label{eq:32}
  h_0 = \kappa\,\Nseg^{1/4}\,\sqrt{\frac{\Sn}{\Tdata}}\,,\quad\text{with}\quad\kappa\sim7-9\,,
\end{equation}
for the same confidence and false-alarm level as Eq.~(\ref{eq:40}), and where $\Nseg$ denotes the number of
semi-coherent segments.

These latter results suggested the inaccurate idea that the sensitivity of semi-coherent searches
follows an exact $\Nseg^{1/4}$ scaling. However, this was later shown \cite{prix2012optimal,wette2012estimating} to not
be generally a good approximation except asymptotically in the limit of a large number of segments ($\Nseg\gtrsim
100-1000)$.

Furthermore, these past sensitivity estimates relied on the assumption of a ``constant signal-to-noise ratio (SNR)'' population of signals.
While this approximation substantially simplifies the problem, it introduces a noticeable bias into the estimate, as
discussed in more detail in \cite{wette2012estimating}.
For example, the constant-SNR bias combined with the incorrect $\Nseg^{1/4}$ scaling in Eq.~\ref{eq:32} would result in
an overestimate by a \emph{factor of two} of the sensitivity of the first Einstein@Home search on LIGO S5 data \cite{S5R5}.

These limitations of previous sensitivity estimates were eventually overcome by the analytic sensitivity-estimation
method developed by \citet{wette2012estimating} for semi-coherent StackSlide $\FF$-statistic searches.
In this work we simplify and extend this framework by employing a simpler direct numerical implementation.
This further improves the estimation accuracy by requiring fewer approximations. It also allows us to generalize the framework to
multi-stage hierarchical \StackSlideF{} searches, \HoughF{} searches (such as \cite{S5R5}), as well as to Bayesian upper
limits based on $\FF$-statistic searches.

\subsection*{Plan of this paper}
\label{sec:plan-this-paper}

  Sec.~\ref{sec:searches} provides a description of the CW signal model and introduces different $\FF$-statistic-based
  search methods.
  In Sec.~\ref{sec:alt-notat-sens} we present the sensitivity-estimation framework and its implementation, for both
  frequentist and Bayesian upper limits.
  Section \ref{sec:upperLimits} discusses how (frequentist) upper limits are typically measured using Monte-Carlo
  injection-recovery simulations.
  Section~\ref{sec:accuracy} provides comparisons of our sensitivity estimates to simulated upper limits in Gaussian
  noise, while in
  Sec.~\ref{ssec: CompEstSim} we provide a comprehensive summary of published sensitivities of past CW
  searches (translated into sensitivity depth), and a comparison to our sensitivity estimates where applicable.
  We summarize and discuss the results in Sec.~\ref{sec:discussion}.
  Further details on the referenced searches and upper limits are given in appendix \ref{sec:SearchOverview}.
  More technical details on the signal model can be found in appendix \ref{sec:cw-signal-model}.
  Finally, appendix~\ref{sec:distr-ff-stat} contains a discussion of the distribution of the maximum $\FF$-statistic
  over correlated templates.

\section{\label{sec:searches}$\FF$-statistic-based search methods}

This section provides an overview of the $\FF$-statistic-based search methods for which sensitivity estimates are
derived in Sec.~\ref{sec:alt-notat-sens}. Further technical details about the signal model and the $\FF$-statistic are given in
appendix \ref{sec:cw-signal-model}.
For a broader review of the CW signal model, assumptions and search methods, see for example \cite{prix2009gravitational,2015PASA...32...34L,2017MPLA...3230035R}

\subsection{Signal model}
\label{sec:signal-model}

For the purpose of sensitivity estimation we assume the data timeseries $x^X(t)$ from each detector $X$ to
be described by Gaussian noise, i.e.\ $n^X(t)\sim\Gauss(0,\Sn^X)$ with zero mean and (single-sided) power-spectral
density (PSD) $\Sn^X$.
A gravitational-wave signal creates an additional strain $h^X(t)$ in the detector, resulting in a timeseries
\begin{equation}
  \label{eq:35}
  x^X(t) = n^X(t) + h^X(t)\,.
\end{equation}
For continuous gravitational waves the two polarization components can be written as
\begin{equation}
  \label{eq:39}
  \begin{split}
    h\plus(\tau) = A\plus\,\cos\left(\phi(\tau)+\phi_0\right)\,,\\
    h\cross(\tau)= A\cross\,\sin\left(\phi(\tau)+\phi_0\right)\,,
  \end{split}
\end{equation}
where $\phi(\tau)$ describes the phase evolution of the signal in the source frame.
For the typical quasi-periodic signals expected from rotating neutron stars, this can be expressed as a Taylor series
expansion around a chosen reference time (here $\tau_{\mathrm{ref}}=0$ for simplicity) as
\begin{equation}
  \label{eq:41}
  \phi(\tau) = 2\pi( f\,\tau + \frac{1}{2}\,\dot{f}\,\tau^2 + \ldots )\,,
\end{equation}
in terms of derivatives of the slowly-varying intrinsic CW frequency $f(\tau)$.
For a triaxial neutron star spinning about a principal axis, the two polarization amplitudes are given by
\begin{equation}
  \label{eq:31}
  A\plus  = \frac{1}{2} h_0\,(1 + \cos^2\iota)\,,\quad
  A\cross = h_0\,\cos\iota\,,
\end{equation}
in terms of the angle $\iota$ between the line of sight and the neutron star rotation axis and the overall signal
amplitude $h_0$. This definition uses the common gauge condition of $A\plus \ge |A\cross|$.
After translating the source-frame signal into the detector frame (see appendix \ref{sec:cw-signal-model} for details),
the strain signal $h^X(t)$ at each detector $X$ can be expressed in the factored form
\begin{equation}
  \label{eq:36}
  h^X(t; \Amp,\lambda) = \sum_{\mu=1}^4 \Amp^\mu\,h^X_\mu(t; \lambda)\,,
\end{equation}
which was first shown in \cite{JKS1998data}, and where the four amplitudes $\Amp^\mu$ depend on the \emph{amplitude
  parameters} $\{h_0,\cos\iota,\psi,\phi_0\}$ as given in Eq.~\ref{eq:37}).
The four basis functions $h_\mu^X(t;\lambda)$, which are given explicitly in Eq.~\eqref{eq:44}, depend on the
\emph{phase-evolution parameters} $\lambda = \{\vSky, f, \dot{f}, \ldots\}$, namely sky position $\vSky$, frequency $f$
and its derivatives $\left.f^{(k)}=d^kf/d\tau^k\right|_{\tau_{\mathrm{ref}}}$, and binary-orbital parameters in the case of a neutron
star in a binary.

\subsection{Coherent $\FF$-statistic}
\label{ssec: methods}

For pure Gaussian-noise timeseries $\{n^X(t)\}$ in all detectors $X$, the likelihood can be written as
(e.g.\ see\cite{finn92:_detection,MultiDetFstat,prix2007search}):
\begin{equation}
  \label{eq:46}
  \prob{x=n}{\Sn} = \kappa\,e^{-\frac{1}{2}\,\scalar{n}{n}}\,,
\end{equation}
in terms of the multi-detector scalar product
\begin{equation} \label{eq:scalP}
  \scalar{x}{y} \equiv 4 \Re \sum_X \int_0^\infty \frac{\tilde x^X(f)\,\tilde y^{X*}(f)}{\Sn^X(f)}\DD f\,,
\end{equation}
where $\tilde{x}(f)$ denotes the Fourier transform of $x(t)$, and $x^*$ denotes complex conjugation of $x$.
Using the additivity of noise and signals (cf.\ Eq.~\eqref{eq:35}), we can express the likelihood for data $x$,
assuming Gaussian noise plus a signal $h(\Amp,\lambda)$ as
\begin{align}
  \prob{x}{\Sn, \Amp,\lambda} &= \prob{x - h(\Amp,\lambda)}{\Sn} \notag\\
                              &= \kappa \, e^{-\frac{1}{2}\scalar{(x-h)}{(x-h)}}\,.  \label{eq:48}
\end{align}
From this we obtain the log-likelihood ratio between the signal and noise hypotheses as
\begin{align}
  \label{eq:49}
  \ln \Lambda(x; \Amp, \lambda) &\equiv \ln\frac{\prob{x}{\Sn, \Amp,\lambda}}{\prob{x}{\Sn}} \notag\\
                                & = \scalar{x}{h} - \frac{1}{2}\scalar{h}{h}\,.
\end{align}
Analytically maximizing the log-likelihood ratio over $\Amp$ (c.f.\ appendix \ref{sec:cw-signal-model}) yields the
$\FF$-statistic \cite{JKS1998data}:
\begin{equation}
  \label{eq:58}
  \FF(x;\lambda) \equiv \max_{\Amp} \ln \Lambda(x;\Amp, \lambda)
\end{equation}
The statistic $2\FF$ follows a $\chi^2$-distribution with four degrees of freedom and non-centrality $\rho^2$,
\begin{equation}
  \label{eq:13}
  \prob{2\FF}{\rho^2} = \chi^2_4(2\FF; \rho^2)\,,
\end{equation}
with expectation and variance
\begin{equation}
  E[2\FF] = 4 + \rho^2\,,\quad \mathrm{var}[2\FF] = 8 + 4\rho^2\,,
\end{equation}
where $\rho$ corresponds to the signal-to-noise ratio (SNR) of coherent matched filtering.

In the \emph{perfect-match} case, where the template phase-evolution parameters $\lambda$ coincide
with the parameters $\lambda\signal$ of a signal in the data $x$, the SNR can be explicitly expressed as
\begin{equation}
  \label{eq:1}
  \rhoOpt^2 \equiv \scalar{h}{h} = \frac{4}{25}\,\frac{h_0^2}{\Sn}\,\Tdata\,R^2(\theta)\,,
\end{equation}
where $\Tdata$ is the total amount (measured as time) of data over all detectors, and $\Sn$ denotes the multi-detector
noise floor, defined as the \emph{harmonic mean} over the per-detector PSDs $\Sn^X$, namely
\begin{equation}
  \label{eq:52}
  \frac{1}{\Sn} \equiv \frac{1}{N}\,\sum_X \frac{1}{\Sn^X}\,.
\end{equation}
Note that in practice the $\FF$-statistic-based search implementations do not assume stationary noise over the whole
observation time, but only over short durations of order $\TSFT\sim\SI{30}{mins}$, corresponding to the length of the
Short Fourier Transforms (SFTs) that are typically used as input data. The present formalism can straightforwardly be
extended to this case \cite{prix:_cfsv2}, where the relevant overall multi-detector noise-PSD definition $\Sn$ generalizes
as the \emph{harmonic mean over all SFTs}, namely
\begin{equation}
  \label{eq:23}
  \frac{1}{\Sn} \equiv \frac{1}{\NSFT}\sum_{\alpha} \frac{1}{\Sn^{\alpha}}\,,
\end{equation}
where $\alpha$ is an index enumerating all SFTs (over all detectors), and $\Sn^\alpha$ is the corresponding noise PSD
estimated for SFT $\alpha$.

The response function $R(\theta)$ (following the definition in \cite{wette2012estimating}) depends on the subset of
signal parameters
\begin{equation}
  \label{eq:7}
  \theta\equiv\{\vSky, \cos\iota, \psi\}\,,
\end{equation}
and can be explicitly expressed \cite{prix2010f} as
\begin{equation}
  \label{eq:2}
  R^2(\theta) = \frac{25}{4}\left[
    \al_1\,A(\vSky) + \al_2\,B(\vSky) + 2\al_3\,C(\vSky)
  \right]\,,
\end{equation}
with the sky-dependent antenna-pattern coefficients $\{A,B,C\}$ of Eq.~\eqref{eq:51}, and
\begin{align}
  \label{eq:3}
  \al_1 &\equiv \frac{1}{4}\,(1+\cos^2\iota)^2\,\cos^22\psi + \cos^2\iota\,\sin^22\psi\,,\\
  \al_2 &\equiv \frac{1}{4}\,(1+\cos^2\iota)^2\,\sin^22\psi + \cos^2\iota\,\cos^22\psi\,,\\
  \al_3 &\equiv \frac{1}{4}\,(1-\cos^2\iota)^2\,\sin2\psi\cos2\psi\,.
\end{align}
One can show that $R^2$ averaged over $\psi\in[-\pi/4,\pi/4]$ and $\cos\iota\in[-1,1]$
yields
\begin{equation}
  \label{eq:4}
  \avg{R^2}_{\cos\iota,\psi} = \frac{5}{2}\left(A(\vSky) + B(\vSky)\right)\,,
\end{equation}
and further averaging $\vSky$ isotropically over the sky yields
\begin{equation}
  \label{eq:5}
  \avg{R^2}_{\theta} = 1\,.
\end{equation}
Using this with Eq.~\eqref{eq:1} we can therefore recover the sky- and polarization-averaged squared-SNR expression
(e.g.\ see \cite{JKS1998data}):
\begin{equation}
  \label{eq:6}
  \avg{\rhoOpt^2}_{\theta} = \frac{4}{25}\frac{h_0^2}{\Sn}\,\Tdata\,.
\end{equation}

\subsection{\label{ssc: semi-coherent} Semi-coherent $\FF$-statistic methods}
Semi-coherent methods \cite{2000PhRvD..61h2001B} typically divide the data into $\Nseg$ shorter segments of duration
$\Tseg < T_{\mathrm{obs}}$. The segments are analyzed coherently, and the per-segment detection statistics are
combined incoherently. Generally this yields lower sensitivity for the same amount of data analyzed than a
fully-coherent search. However, the computational cost for a fully-coherent search over the same amount of
data is often impossibly large, while the semi-coherent cost can be tuned to be affordable and typically ends up being
more sensitive at fixed computing cost \cite{2000PhRvD..61h2001B,2005PhRvD..72d2004C,prix2012optimal}.

There are a number of different semi-coherent methods currently in use, such as PowerFlux, FrequencyHough, SkyHough,
TwoSpect, CrossCorr, Viterbi, Sideband, loosely-coherent statistics and others (e.g.\ see \cite{2017MPLA...3230035R} and references therein).
Many of these methods work on short segments, typically of length $\Tseg\sim\SI{30}{min}$, and use Fourier power in
the frequency bins of these Short Fourier Transforms (SFTs) as the coherent base statistic.

In this work we focus exclusively on sensitivity estimation of $\FF$-statistic-based methods, namely \StackSlideF{}
(e.g.\ see \cite{prix2012optimal}) and \HoughF{} introduced in \cite{krishnan2004hough}. Here the length of segments is
only constrained by the available computing cost, and segments will typically span many hours to days, which yields
better sensitivity, but also requires higher computational cost.
Therefore, many of the computationally expensive semi-coherent $\FF$-statistic searches are run on the distributed
Einstein@Home computing platform \cite{EinsteinAtHome}.

Note that these methods are not to be confused with the (albeit closely related) ``classical'' StackSlide and Hough
methods, which use SFTs as coherent segments, as described for example in \cite{S4AS}.

\subsubsection{\StackSlideF{}: summing $\FF$-statistics}
\label{sec:stacksl-summ-ff}

The \StackSlideF{} method uses the sum of the coherent per-segment $\FCO$-statistic values in a given
parameter-space point $\lambda$ as the detection statistic, namely
\begin{equation}
  \label{eq:11}
  2\FSC \equiv \sum_{\ell = 1}^{\Nseg} 2\FCO_\ell\,,
\end{equation}
where $\FCO_\ell$ is the coherent $\FF$-statistic of Eq.~\eqref{eq:58} in segment $\ell$.
This statistic follows a $\chi^2$-distribution with $4\Nseg$ degrees of freedom
and non-centrality $\rho^2$, i.e.\
\begin{equation}
  \label{eq:12}
  \prob{2\sumFstat}{\rho^2} = \chi^2_{4\Nseg}(2\sumFstat;\rho^2)\,,
\end{equation}
where the {non-centrality} $\rho^2$ is identical to the expression for the \emph{coherent squared SNR} of Eq.~\eqref{eq:1},
with $\Tdata$ referring to the whole data set used, and $\Sn$ is the corresponding noise floor.
However, the non-centrality in the semi-coherent case cannot be considered a ``signal to noise ratio'', due to the
larger $\Nseg$-dependent degrees of freedom of the $\chi^2$ distribution compared to Eq.~\eqref{eq:13}, which increases
the false-alarm level at fixed threshold and reduces the ``effective'' semi-coherent $\SC{\text{SNR}}$ to
$\SC{\text{SNR}}^2=\rho^2/\sqrt{\Nseg}$ (e.g.\ see [Eq.(14)] in \cite{2015PhRvD..92h2003W}).

The expectation and variance for $2\sumFstat$ are
\begin{equation}
  E[2\FF] = 4\Nseg + \rho^2\,,\quad \mathrm{var}[2\FF] = 8\Nseg + 4\rho^2\,.
\end{equation}
We note that in practice \StackSlideF{} searches often quote the average $\avgFstat$ over segments instead of the sum
$\FSC$, i.e.\
\begin{equation}
  \avgFstat \equiv \frac{1}{\Nseg} \FSC\,.
\end{equation}


\subsubsection{\HoughF{}: summing threshold crossings}
\label{sec:hough-ff:-summing}

The \HoughF{} method \cite{krishnan2004hough} sets a threshold $\FCO\thr$ on the per-segment coherent
$\FCO$-statistics and uses the number of threshold-crossings over segments as the detection
statistic, the so-called \emph{Hough number count} $\nc$, i.e.\
\begin{equation}
  \label{eq:24}
  \nc \equiv \sum_{\ell=1}^{\Nseg} \Theta(\FCO_\ell-\FCO\thr)\,,
\end{equation}
where $\Theta(x)$ is the Heaviside step function.

\subsection{Mismatch and template banks}
\label{sec:templ-banks-mism}

In wide-parameter-space searches the unknown signal parameters $\lambda \in \parSpace$ are assumed to fall somewhere
within a given search space $\parSpace$.
In this case one needs to compute a statistic (such as those defined in the previous sections) over a whole ``bank'' of
templates $\templBank\equiv\{\lambda_i\}_{i=1}^{\Ntempl}$. This template bank has to be chosen in such a way that
any putative signal $\lambda\signal\in\parSpace$ would suffer only an acceptable level of loss of SNR.
This is typically quantified in terms of the so-called \emph{mismatch} $\mis$, defined as the relative loss of
$\rho^2(\lambda\signal;\lambda)$ in a template $\lambda$ with respect to the perfect-match
$\rho^2(\lambda\signal;\lambda\signal) = \rhoOpt^2$ (of Eq.~\eqref{eq:1}), namely
\begin{equation}
  \label{eq:54}
  \mis(\lambda\signal;\lambda) \equiv \frac{\rho^2(\lambda\signal;\lambda\signal) - \rho^2(\lambda\signal;\lambda)}{\rho^2(\lambda\signal;\lambda\signal)}\,.
\end{equation}
We can therefore express the ``effective'' non-centrality parameter $\rhoEff^2$ in a template point $\lambda$ in the
$\FF$-statistic $\chi^2$-distribution of Eqs.~\eqref{eq:13},\eqref{eq:12} as
\begin{equation}
  \label{eq:55}
  \rhoEff^2 \equiv \rho^2(\lambda\signal;\lambda) = \left( 1 - \mis\right)\,\rhoOpt^2\,.
\end{equation}

\subsection{Sensitivity Depth}
\label{sec:sensitivity-depth}

The $\FF$-statistic non-centrality parameter $\rho^2$ depends on signal amplitude $h_0$ and overall noise floor $\Sn$
(cf.\ Eq.~\eqref{eq:23}) only through the combination $h_0/\sqrt{\Sn}$, as seen in Eq.~\eqref{eq:1}.
The sensitivity of a search is therefore most naturally characterized in terms of the so-called \emph{sensitivity depth}
\cite{behnke2015postprocessing}, defined as
\begin{equation}
  \label{eq:10}
  \depth \equiv \frac{\sqrt{\Sn}}{h_0}\,,
\end{equation}
in terms of the overall noise PSD $\Sn$ defined as the harmonic mean over all SFTs used in the search, see
Eq.~\eqref{eq:23}.

A particular choice of search parameters ($\Nseg,\Tdata$, template bank) will generally yield a frequency-dependent
upper limit $h_0(f)$, due to the frequency-dependent noise floor $\Sn(f)$.
However, for fixed search parameters this will correspond to a \emph{constant} sensitivity depth $\depth$, which is
therefore often a more practical and natural way to characterize the performance of a search, independently of
the noise floor.

\section{Sensitivity Estimate}
\label{sec:alt-notat-sens}

As discussed in more detail in the introduction, by \emph{sensitivity} we mean the (measured or expected) upper limit on
$h_0$ for a given search (or equivalently, the sensitivity depth $\depth=\sqrt{\Sn}/h_0$), which can either refer to the
frequentist or Bayesian upper limit.

\subsection{Frequentist upper limits}
\label{sec:estim-freq-upper}

The frequentist upper limit is defined as the weakest signal amplitude $h_0$ that can be detected at a given detection
probability $\pDET$\footnote{or equivalently, false-dismissal probability $\pFD = 1 - \pDET$} (typically chosen as
$90\%$ or $95\%$) above a threshold $\stat\thr$ on a statistic $\stat(x)$.
The threshold can be chosen as the loudest candidate obtained in the search, or it can be set corresponding to a
desired false-alarm level $\pFA$ (or p-value), defined as
\begin{equation}
  \pFA(\stat\thr) \equiv \prob{\stat>\stat\thr}{h_0=0}\,,   \label{eq:8a}
\end{equation}
which can be inverted to yield $\stat\thr = \stat\thr(\pFA)$.
The detection probability for signals of amplitude $h_0$ is
\begin{equation}
  \pDET(\stat\thr; h_0) \equiv \prob{\stat>\stat\thr}{h_0}\,,   \label{eq:8b}
\end{equation}
which can be inverted to yield the upper limit $h_0(\stat\thr,\pDET)$.

We can write $\pFA(\stat\thr) = \pDET(\stat\thr; h_0=0)$, and so we can express both in terms of the general
threshold-crossing probability as
\begin{equation}
  \label{eq:9}
  \prob{\stat>\stat\thr}{h_0} = \int_{\stat\thr}^{\infty}\prob{\stat}{h_0} \DD\stat\,.
\end{equation}

\subsection{Approximating wide-parameter-space statistics}
\label{sec:appr-wide-param}

As discussed in Sec.~\ref{sec:templ-banks-mism}, a wide parameter-space search for unknown signals
$\lambda \in \parSpace$ typically proceeds by computing a (single-template) statistic over a bank of templates
$\templBank\equiv\{\lambda_i\}_{i=1}^{\Ntempl}$ covering the parameter space $\parSpace$.
This results in a corresponding set of (single-template) statistic values $\{\stat\single(x;\lambda_i)\}$, which need to
be combined to form the overall wide-parameter-space statistic $\stat(x)$.
This would naturally be obtained via marginalization (i.e.\ integrating the likelihood over $\parSpace$), but in
practice is mostly done by maximizing the single-template statistic over $\templBank$, i.e.
\begin{equation}
  \label{eq:30}
  \stat(x) \equiv \stat\loudest(x) \equiv \max_{\lambda_i\in\templBank}\stat\single(x;\lambda_i)\,.
\end{equation}

\subsubsection{Noise case: estimating the p-value $\pFA$}
\label{sec:noise-case:-estim}

For the pure noise case of Eq.~\eqref{eq:8a}, it is difficult to determine a reliable expression for
$\prob{\stat\loudest}{h_0=0}$, even if the single-template statistic $\prob{\stat\single}{h_0=0}$ follows a known
distribution (such as for the $\FF$-based statistics discussed in Sec.~\ref{sec:searches}).
The reason for this difficulty lies in the correlations that generally exist between ``nearby'' templates in
$\lambda_i\in\templBank$.

If all $\Ntempl$ templates were strictly uncorrelated, one could use the well-known
expression Eq.~\eqref{eq:56} \cite{S5CasA,wette2012estimating} for the distribution of the maximum.
In this case one can also relate the single-trial p-value $\pFA\single \approx \pFA/\Ntempl$ to the
wide-parameter-space p-value $\pFA$ (for $\pFA\single\ll1$).

Although it is a common assumption in the literature, template correlations do not simply modify the ``effective'' number
of independent templates to use in Eq.~\eqref{eq:56}, but they generally also affect the functional form of the
underlying distribution for the maximum $\stat\loudest$, as illustrated in appendix \ref{sec:distr-ff-stat} with a
simple toy model.

In this work we assume that the upper limit refers to a known detection threshold in Eq.~\eqref{eq:8b}.
This can be obtained either from (i) the loudest observed candidate (the most common situation in real searches), or from
(ii) setting a single-template p-value $\pFA\single$ and inverting the known single-template distribution Eq.~\eqref{eq:8a},
or from (iii) a numerically-obtained relation between $\pFA$ and the threshold $\stat\thr$, e.g.\ via Monte-Carlo
simulation.

\subsubsection{Signal case: estimating the detection probability $\pDET$}
\label{sec:sign-case:-estim}

In the signal case it is easier to estimate the maximum-likelihood statistic $\stat\loudest(x)$ over the full
template bank $\templBank$, provided we can assume that the highest value of $\stat\single$ will be realized near the
signal location, which should be true as long as the p-value $\pFA$ is low (typically $\pFA\lesssim1\%$)
and the signals have relatively high detection probability (typically $\pDET\sim 90\%$ or $95\%$).
This will typically be a good approximation, but in Sec.~\ref{sec:accuracy} we will also encounter situations where
deviations from the predictions can be traced to violations of these assumptions.
We therefore approximate
\begin{equation}
  \label{eq:33}
  \stat\loudest(x) \approx \stat\single(x;\lambda\loudest)\,,
\end{equation}
where $\lambda\loudest$ is the ``closest'' template $\in\templBank$ to the signal location $\lambda\signal$,
defined in terms of the metric Eq.~\eqref{eq:54}, namely the template with the smallest mismatch $\mis$ from the
signal. This template yields the highest effective non-centrality parameter over the template bank, namely
\begin{equation}
  \label{eq:34}
  \rhoEff^2 \equiv \rho^2(\lambda\signal;\lambda\loudest) = (1-\mis)\,\rhoOpt^2(\lambda\signal)\,.
\end{equation}

\subsection{\StackSlideF{} sensitivity}
\label{sec:single-stage-stacksl}

We first consider a semi-coherent \StackSlideF{} search using the summed $\sumFstat$-statistic of
Eq.~\eqref{eq:11}, i.e.\ $\stat\single(x;\lambda) = 2\FSC(x;\lambda)$.
This case also includes fully-coherent $\FF$-statistic searches, which simply correspond to the special case $\Nseg=1$.

We see from Eq.~\eqref{eq:9} that in order to estimate the sensitivity, we need to know $\prob{2\FSC}{h_0}$.
This can be obtained via marginalization (at fixed $h_0$) of the known distribution $\prob{2\FSC}{\rho^2}$ of
Eq.~\eqref{eq:12}, combined with the assumption Eq.~\eqref{eq:34} that the highest statistic value will occur in the
``closest'' template, with mismatch distribution $\prior{\misSC}$:
\begin{align}
  \label{eq:14}
  \prob{2\FSC}{h_0} &= \int \prob{2\FSC,\theta,\misSC}{h_0}\,\DD^4\theta \DD\misSC \notag\\
  &= \int \prob{2\FSC}{h_0,\,\theta,\,\misSC}\prior{\theta}\prior{\misSC}\,\DD^4\theta  \DD\misSC \notag\\
  &= \int \prob{2\FSC}{\rhoEff^2}\,\prior{\theta}\,\prior{\misSC}\,\DD^4\theta \DD\misSC\,,
\end{align}
where $\rhoEff^2(h_0,\theta,\misSC) = \rhoOpt^2(h_0,\theta)\,(1-\misSC)$ in terms of the perfect-match non-centrality $\rhoOpt^2$
defined in Eq.~\eqref{eq:1}, and in the last step we used the fact that the distribution for $2\FSC$ is fully specified
in terms of the non-centrality parameter $\rho^2$ of the $\chi^2$-distribution with $4\Nseg$ degrees of freedom, as given
in Eq.~\eqref{eq:12}.

Equation ~\eqref{eq:14} requires five-dimensional integration for each sensitivity estimation, which would be slow and
cumbersome.
One of the key insights in \cite{wette2012estimating} was to notice that the perfect-match SNR $\rhoOpt$ of Eq.~\eqref{eq:1}
depends on the four parameters $\theta$ only through the scalar $R^2(\theta)$, and we can therefore use a reparametrization
\begin{equation}
  \label{eq:16}
  \int_{\theta(R^2)}\prior{\theta}\,\DD^4\theta = \prior{R^2}\,\DD R^2\,,
\end{equation}
where $\theta(R^2)$ denotes the subspace of $\theta$ values yielding a particular $R^2$ from Eq.~\eqref{eq:2}.

The one-dimensional distribution $\prior{R^2}$ can be obtained by Monte-Carlo sampling over the priors of
sky-position $\vSky$ (typically either isotropically over the whole sky, or a single sky-position in case of a directed
search) and polarization angles $\cos\iota$ (uniform in $[-1,1]$) and $\psi$ (uniform in $[-\pi/4,\pi/4]$). The
resulting values of $R^2(\theta)$ are histogrammed and used as an approximation for $P(R^2)$, which can be reused for
repeated sensitivity estimations with the same $\theta$-priors.
We can therefore rewrite Eq.~\eqref{eq:14} as
\begin{equation}
  \label{eq:17}
  \prob{2\FSC}{h_0} = \int \prob{2\FSC}{\rhoEff^2}\prior{R^2}\prior{\misSC}\DD R^2 \DD\misSC\,,
\end{equation}
with
\begin{align}
  \prob{2\FSC}{\rhoEff^2} &= \chi^2_{4\Nseg}(2\FSC; \rhoEff^2)\,,  \label{eq:18b}\\
  \rhoEff^2(h_0,R^2,\misSC) &= \frac{4}{25}\,\frac{h_0^2}{\Sn}\,\Tdata\,R^2\,(1-\misSC)\,.   \label{eq:18a}
\end{align}
The mismatch distribution $\prior{\mis}$ for any given search will typically be obtained via injection-recovery
Monte-Carlo simulation, where signals are repeatedly generated (without noise) and searched for over the template bank,
obtaining the corresponding mismatch $\misSC$ for each injection. This is often a common step in validating a search
and template-bank setup. Alternatively, for some search methods pre-computed estimates for the mismatch distributions
exist as a function of the template-bank parameters, e.g.\ for the \textsc{Weave} search code \cite{2018arXiv180403392W}.

Inserting Eq.~\eqref{eq:17} into the detection probability of Eq.~\eqref{eq:9}, we obtain
\begin{equation}
  \label{eq:19}
  \pDET(2\FSC\thr; h_0) = \int \pDET(2\FSC\thr; \rhoEff^2) P(R^2)P(\misSC) \DD R^2 \DD\misSC\,,
\end{equation}
where
\begin{equation}
  \label{eq:20}
  \pDET(2\FSC\thr; \rhoEff^2) \equiv \int_{2\FSC\thr}^\infty \chi^2_{4\Nseg}(2\FSC; \rhoEff^2)\, \DD2\FSC\,.
\end{equation}
Equation \eqref{eq:19} can be easily and efficiently computed numerically, and simple inversion (via 1-D root-finding)
yields the sensitivity (i.e.\ upper limit) $h_0$ for given detection probability $\pDET$ and threshold $2\FSC\thr$.

\subsection{Multi-stage \StackSlideF{} sensitivity}
\label{sec:multi-stage-stacksl}

The sensitivity estimate for a single \StackSlideF{} search can be generalized to
hierarchical multi-stage searches, where threshold-crossing candidates of one search stage are followed up by deeper
subsequent searches in order to increase the overall sensitivity (e.g.\ see
\cite{2000PhRvD..61h2001B,2005PhRvD..72d2004C,ShaltevPrix2013:_coherent_followup,S6BucketFU,O1ASE@H}).
We denote the $\nStages$  stages with an index $i = 1 \ldots \nStages$.
Each stage $i$ is characterized by the number $\Nseg\stage{i}$ of segments, the amount of data $\Tdata\stage{i}$, the
noise PSD $\Sn\stage{i}$, a mismatch distribution $\prior{\misSC\stage{i}}$, and a threshold $2\FSC\thr\stage{i}$
(corresponding to a false-alarm level $\pFA\stage{i}$ at that stage).

The initial wide-parameter-space search (stage $i=1$) yields candidates that cross the threshold $2\Fth\stage{1}$ in
certain templates $\{\lambda\}$. The next stage follows up these candidates with a more sensitive search,
which can be achieved by reducing the mismatch $\misSC\stage{i}$ (choosing a finer template bank grid), or by increasing
the coherent segment length (and reducing the number of segments $\Nseg\stage{i}$).
Often the final stage $i=\nStages$ in such a follow-up hierarchy would be fully coherent, i.e.\ $\Nseg\stage{\nStages}=1$.

In order for any given candidate (which can be either due to noise or a signal) to cross the final threshold
$2\FF^{(\nStages)}$, it has to cross \emph{all} previous thresholds as well, in other words Eq~\eqref{eq:8a},\eqref{eq:8b} now generalize to
\begin{equation}
  \label{eq:21}
  \pDET\total(h_0) = \prob{ \{2\FSC\stage{i}>2\FSC\thr\stage{i}\}_{i=1}^{\nStages}}{h_0}\,.
\end{equation}
In order to make progress at this point we need to assume that the threshold-crossing probabilities in different stages
are \emph{independent} of each other, so for $j\not=i$ we assume
\begin{equation}
  \label{eq:8}
  \prob{2\FSC\stage{i}\!>\!2\FSC\thr\stage{i}}{\rho^2, 2\FSC\stage{j}\!>\!2\FSC\thr\stage{j}} = \prob{2\FSC\stage{i}\!>\!2\FSC\thr\stage{i}}{\rho^2}\,,
\end{equation}
which would be exactly true if the different stages used different data (see also \cite{2005PhRvD..72d2004C}).
In the case where the same data is used in different stages, this approximation corresponds to an \emph{uninformative}
approach, in the sense that we do not know how to quantify and take into account the correlations between the statistics
in different stages. We proceed without using this potential information, which could in principle be used to improve
the estimate.
It is not clear if and how much of an overall bias this approximation would introduce. A detailed study of this question
is beyond the scope of this work and will be left for future study.


Using the assumption of independent stages we write
\begin{align}
  \pDET\total(h_0) &= \int \prod_{i = 1}^{\nStages} \pDET\stage{i}(2\FSC\thr\stage{i}; h_0, R^2)\,\prior{R^2} \DD R^2\,,\label{eq:15a}\\
  \pFA\total &= \prod_{i = 1}^{\nStages} \pFA\stage{i}(2\FSC\thr\stage{i})\,,  \label{eq:15b}
\end{align}
where now the $R^2$-marginalization needs to happen over all stages combined, as the signal parameters $R^2(\theta)$ are
intrinsic to the signal and therefore independent of the stage.
On the other hand, the mismatch distribution $P(\misSC\stage{i})$ depends on the stage, as each stage will typically use
a different template grid, and so we have
\begin{equation}
  \label{eq:18}
  \pDET\stage{i}(2\FSC\thr\stage{i}; h_0,R^2) = \int_{0}^{1}\pDET\stage{i}(2\FSC\thr\stage{i};
  \rhoEff^{2\,(i)})\,\prior{\misSC\stage{i}}\DD\misSC\stage{i}\,,
\end{equation}
where $\pDET(2\FSC\thr;\rhoEff^2)$ is given by Eq.~\eqref{eq:20} using the respective per-stage values.

Equation~\eqref{eq:15a} can easily be solved numerically and inverted for the sensitivity $h_0$ at given $\pDET\total$
and a set of thresholds $\{2\FSC\thr\stage{i}\}$.

Note that in practice one would typically \cite{S6BucketFU} want to choose the thresholds in such a way that a signal
that passed the 1st-stage threshold $2\FSC\thr\stage{1}$ should have a very low probability of being discarded by
subsequent stages, in other words $\pDET\stage{i>1} \approx 1$, and therefore $\pDET\total(h_0)\approx
\pDET\stage{1}(2\FSC\thr\stage{1};h_0)$.
Therefore subsequent stages mostly serve to reduce the total false-alarm level $\pFA\total$,
allowing one to increase the first-stage $\pFA\stage{1}$ by lowering the corresponding threshold $\FSC\stage{1}$,
resulting in an overall increased sensitivity.

\subsection{\HoughF{} sensitivity}
\label{sec:hough-tracking-ff}

Here we apply the sensitivity-estimation framework to the \HoughF{} statistic introduced in Sec.~\ref{sec:hough-ff:-summing}.
The key approximation we use here is to assume that for a given signal $\{h_0,R^2(\theta)\}$, the
coherent per-segment $\FCO_\ell$-statistic has the same threshold-crossing probability $\pTH$ in each segment $\ell$,
i.e.\ $\pTH^\ell=\pTH$ for all $\ell = 1\ldots\Nseg$, and
\begin{align}
  \label{eq:22}
  \pTH^\ell(h_0,R^2) &\equiv \prob{2\FCO_\ell>2\FCO\thr}{h_0,R^2} \notag\\
              &=\pDET^\ell(2\FCO\thr; h_0,R^2) \notag\\
              &=\int_{0}^{1} \pDET(2\FCO\thr; \rhoEffPerSeg^2)\,\prior{\misCO}\DD\misCO\,,
\end{align}
where the per-segment effective SNR $\rhoEffPerSeg$ is given by Eq.~\eqref{eq:18a} with $\Tdata$ and $\Sn$ referring to
the respective per-segment quantities, and $\misCO$ is the mismatch of $\FCO$-statistic in a single segment.

Provided these quantities are reasonable constant across segments, for a fixed signal $\{h_0,R^2\}$ we can write the
probability for the Hough number count $\nc$ of Eq.~\eqref{eq:24} as a binomial distribution, namely
\begin{equation}
  \prob{\nc}{h_0,R^2} = \dbinom{\Nseg}{\nc}\, \pTH^{\nc}\,(1-\pTH)^{\Nseg-\nc}\,,  \label{eq:23b}
\end{equation}
with $\pTH(h_0,R^2)$ given by Eq.~\eqref{eq:22}.
For a given threshold $\ncTH$ on the number count we therefore have the detection probability
\begin{equation}
  \label{eq:27}
  \pDET(\ncTH; h_0, R^2) = \sum_{\nc=\ncTH}^{\Nseg}\prob{\nc}{h_0, R^2}\,,
\end{equation}
and marginalization over $R^2$ yields the corresponding detection probability at fixed amplitude $h_0$, namely
\begin{equation}
  \label{eq:26}
  \pDET(\ncTH; h_0) = \int \pDET(\ncTH; h_0, R^2)\,\prior{R^2}\DD R^2\,,
\end{equation}
We can numerically solve this for $h_0$ at given $\pDET$ and number-count threshold $\ncTH$, which yields the desired
sensitivity estimate.

\subsection{Bayesian Upper Limits}
\label{sec:bayes-sens-estim}

Bayesian upper limits are conceptually quite different \cite{FreqVsBayesUL} from the frequentist ones discussed up to
this point.
A Bayesian upper limit $h_0^C$ of given confidence (or ``credible level'') $C$ corresponds to the interval $[0,h_0^C]$
that contains the true value of $h_0$ with probability $C$.
We can compute this from the posterior distribution $\prob{h_0}{x}$ for the signal-amplitude $h_0$ given data $x$,
namely
\begin{equation}
  \label{eq:29}
  C = \prob{h_0 < h_0^C}{x} = \int_0^{h_0^C} \prob{h_0}{x}\, \DD h_0\,.
\end{equation}
The Bayesian targeted searches (here referred to as \emph{\BayesPE}) for known pulsars (see Table~\ref{tbl: targeted} and
Sec.~\ref{ssec: targeted}) compute the posterior $\prob{h_0}{x}$ directly from the data $x$, using a time-domain method
introduced in \cite{BayesianPE} .

Here we focus instead on $\FF$-statistic-based searches over a template bank.
As discussed in \cite{FreqVsBayesUL}, to a very good approximation we can compute the posterior from the loudest
candidate $2\FF\loudest(x)$ found in such a search, using this as a proxy for the data $x$, i.e.\
\begin{align}
  \label{eq:28}
  \prob{h_0}{x} &\approx \prob{h_0}{2\FF\loudest(x)}\\
  &\propto \prob{2\FF\loudest(x)}{h_0}\,\prior{h_0}\,,
\end{align}
where we used Bayes' theorem, and the proportionality constant is determined by the normalization condition $\int\prob{h_0}{x} \DD h_0 = 1$.

We have already derived the expression for $\prob{2\FF}{h_0}$ in Eq.~\eqref{eq:17}, and for any choice of prior
$\prior{h_0}$ we can therefore easily compute the Bayesian upper limit $h_0^C(2\FF\loudest)$ for given loudest candidate
$2\FF\loudest$ by inverting Eq.~\eqref{eq:29}.

It is common for Bayesian upper limits on the amplitude to choose a uniform (improper) prior in $h_0$ (e.g.\ see
\cite{O1KnownPulsar}), which has the benefit of simplicity, and also puts relatively more weight on larger values of
$h_0$ than might be physically expected (weaker signals should be more likely than stronger ones).
This prior therefore results in larger, i.e.\  ``more conservative'', upper limits than a more physical prior would.

\subsection{Numerical implementation}
\label{sec:numer-impl}

The expressions for the various different sensitivity estimates of the previous sections have been implemented in
\textsc{GNU Octave} \cite{octave}, and are available as part of the \texttt{OctApps} \cite{octapps-JOSS} data-analysis
package for continuous gravitational waves.

The function to estimate (and cache for later reuse) the distribution $\prior{R^2}$ of Eq.~\eqref{eq:16} is
implemented in \texttt{SqrSNRGeometricFactorHist()}.

The sensitivity-depth estimate for \StackSlideF{}-searches is implemented in
\texttt{SensitivityDepthStackSlide()}, both for the single-stage case of Eq.~\eqref{eq:19} and for the general
multi-stage case of Eq.~\eqref{eq:15a}.
For single-stage \StackSlideF{} there is also a function \texttt{DetectionProbabilityStackSlide()} estimating the
detection probability for a given signal depth $\depth$ and detection threshold.

The \HoughF{} sensitivity estimate of Eq.~\eqref{eq:26} is implemented in \texttt{SensitivityDepthHoughF()}.
An earlier version of this function had been used for the theoretical sensitivity comparison in \cite{S5R5} (Sec.~VB,
and also \cite{prixwette:_2012_S5R5_estimate}), where it was found to agree within an rms error of $7\%$ with the
measured upper limits.

The Bayesian $\FF$-based upper limit expression Eq.~\eqref{eq:29} is implemented in \texttt{SensitivityDepthBayesian()}.

Typical input parameters are the number of segments $\Nseg$, the total amount of data $\Tdata$, the mismatch
distribution $\prior{\mis}$, name of detectors used, single-template false-alarm level $\pFA\single$ (or alternatively,
the $\FF$-statistic threshold), and the confidence level $\pDET$. The default prior on sky-position is isotropic
(suitable for an all-sky search), but this can be restricted to any sky-region (suitable for directed or targeted searches).

The typical runtime on a 3GHz Intel Xeon E3 for a sensitivity estimate including computing $\prior{R^2}$ (which is the
most expensive part) is about \SI{25}{seconds} per detector.
When reusing the same $\theta$-prior on subsequent calls, a cached $\prior{R^2}$ is used and the runtime is reduced to
about \SI{10}{seconds} total, independently of the number of detectors used.

\section{\label{sec:upperLimits} Determining Frequentist Upper Limits}

In order to determine the frequentist upper limit (UL) on the signal amplitude $h_0$ defined in Eq.~\eqref{eq:8b}, one
needs to quantify the probability that a putative signal with fixed amplitude $h_0$ (and all other signal parameters drawn
randomly from their priors) would produce a statistic value exceeding the threshold (corresponding to a
certain false-alarm level, or p-value).
The upper limit on $h_0$ is then defined as the value $h_0^{\pDET}$ for which the detection probability is exactly
$\pDET$, typically chosen as $90\%$ or $95\%$, which is often referred to as the \emph{confidence level} of the UL.

Note that here and in the following it will often be convenient to use the sensitivity depth $\depth\equiv\sqrt{\Sn}/h_0$
introduced in Sec.~\ref{sec:sensitivity-depth} instead of the amplitude $h_0$. We denote $\depth^{\pDET}$ as the
sensitivity depth corresponding to the upper limit $h_0^{\pDET}$ (note that this corresponds to a \emph{lower limit}
on depth).

The UL procedure is typically implemented via a Monte-Carlo injection-and-recovery method: a signal of
fixed amplitude $h_0=\sqrt{\Sn}/\depth$ and randomly-drawn remaining parameters is generated in software and added to
the data (either to real detector data or to simulated Gaussian noise). This step is referred to as a \emph{signal injection}.
A search is then performed on this data, and the loudest statistic value $\FF\loudest$ is recorded and compared
against the detection threshold $\FF\thr$. Repeating this injection and recovery step many times and recording the
fraction of times the threshold is exceeded yields an approximation for $\pDET(\FF\thr;\depth)$.
By repeating this procedure over different $\depth$ values and interpolating one can find $\depth^{\pDET}$ corresponding
to the desired detection probability (and therefore also $h_0^{\pDET}$).

We distinguish in the following between \emph{measured} and \emph{simulated} upper limits:
\begin{itemize}
\item \emph{Measured ULs} refer to the published UL results obtained on real detector data. These typically use an identical
  search procedure for the ULs as in the actual search, often using the loudest candidate (over some range of the
  parameter space) from the original search as the corresponding detection threshold for setting the UL.
  The injections are done in real detector data, and typically the various vetoes, data-cleaning and follow-up procedures
  of the original search will also be applied in the UL procedure.

\item \emph{Simulated ULs} are used in this work to verify the accuracy of the sensitivity estimates. They are
  obtained using injections in simulated Gaussian noise, and searching only a small box in parameter space around the
  injected signal locations.
  The box size is empirically determined to ensure that the loudest signal candidates are always recovered within the
  box. Only the original search statistic is used in the search without any further vetoes or cleaning.
\end{itemize}
A key difference between (most) published (measured) ULs and our simulated ULs concerns the method of
interpolation used to obtain $\depth^{\pDET}$: in practice this is often obtained via a
{sigmoid $\pDET$-interpolation} approach (Sec.~\ref{sec:sigmoid-pdet-fitting}), while we use (and
advocate for) a (piecewise) {linear threshold interpolation} (Sec.~\ref{sec:line-thresh-fitt}) instead.

\subsection{Sigmoid $\pDET$ interpolation}
\label{sec:sigmoid-pdet-fitting}

\begin{figure}[htb]
  \centering
  \includegraphics[width = \columnwidth]{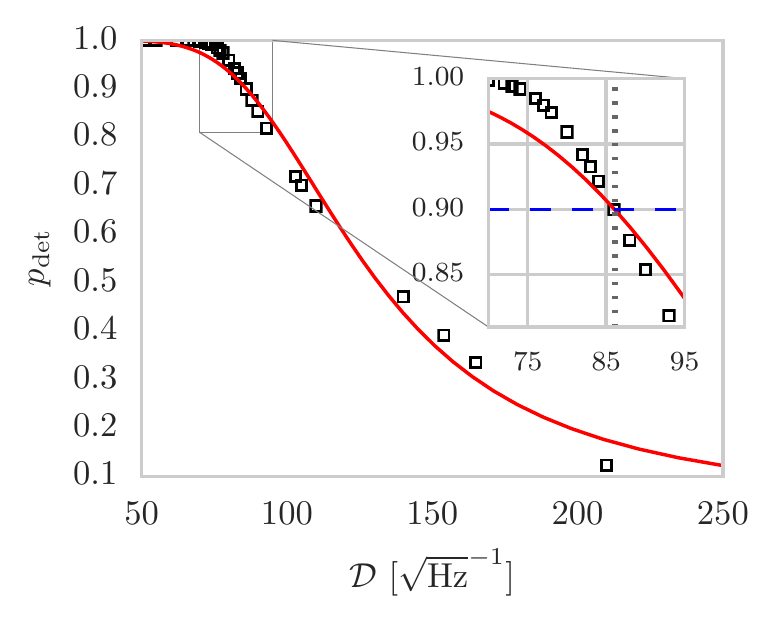}
  \caption{Detection probability $\pDET$ versus sensitivity depth $\depth$ for the \search{S6}{CasA}{\GCT} search
    (cf.~Table~\ref{tbl: directed} and Sec.~\ref{ssec: directed}), using a detection threshold of $2\avgFstat\thr = 8$.
    The squares indicate the results from a simulation in Gaussian noise, while the solid line gives the best-fit
    sigmoid of Eq.~\eqref{eq:47}.
  }
  \label{fig: Sigmoid}
\end{figure}

In this approach one fixes the detection threshold $\FF\thr$ and determines the corresponding $\pDET$ for any given
fixed-$\depth$ injection set. The corresponding functional form of $\pDET(\depth)$ has a qualitative ``sigmoid'' shape
as illustrated in Fig.~\ref{fig: Sigmoid}. An actual sigmoid function of the form
\begin{equation}
  \label{eq:47}
  y(\depth) = \frac{1}{1 + e^{-k\,(\depth-\depth_0)}}\,,
\end{equation}
is then fit to the data by adjusting the free parameters $k$ and $\depth_0$, and from this one can obtain an
interpolation value for $\depth^{\pDET}$.

One problem with this method is that the actual functional form of $\pDET(\depth)$ is not analytically known, and does
not actually seem to be well described by the sigmoid of Eq.~\eqref{eq:47}, as seen in Fig.~\ref{fig: Sigmoid}.
In this particular example the true value at $\pDET=90\%$ just so happens to lie very close to the sigmoid fit, but the
deviation is quite noticeable at $\pDET=95\%$ (see the zoomed inset in Fig.~\ref{fig: Sigmoid}).

Another problem with this method is that the range of depths required to sample the relation $\pDET(\depth)$ often needs
to be quite wide, due to initial uncertainties about where the UL value would be found, which can compound
the above-mentioned sigmoid-fitting problem. Furthermore, the injection-recovery step can be quite computationally
expensive, limiting the number of trials and further increasing the statistical uncertainty on the $\pDET$ measurements.

Both of these problems can be mitigated to some extent by using the sensitivity-estimation method described in this paper
(Sec.~\ref{sec:alt-notat-sens}) to obtain a fairly accurate initial guess about the expected UL value, and then
sample only in a small region around this estimate, in which case even a linear fit would probably yield good accuracy.

\subsection{Piecewise-linear threshold interpolation}
\label{sec:line-thresh-fitt}

\begin{figure}[htb]
  \centering
  \includegraphics[width = \columnwidth]{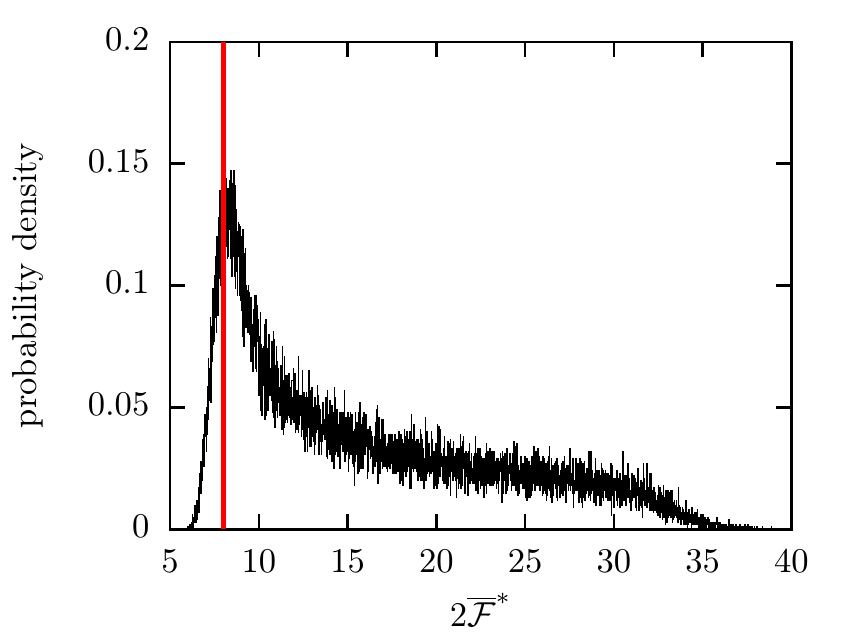}
  \caption{Histogram of recovered loudest $2 \avgFstat$ values for repeated searches on signal injections at fixed
    sensitivity depth $\depth = \SI{86}{\perRootHz}$ (with all other signal parameters randomized), using the search
    setup of the \search{S6}{CasA}{\GCT} directed search.
    The vertical line indicates the resulting threshold value $2\avgFstat\thr = 7.995$ corresponding to $\pDET=90\,\%$
    for this injection set.}
  \label{fig: FindThreshold}
\end{figure}

\begin{figure}[htb]
  \centering
  \includegraphics[width = \columnwidth]{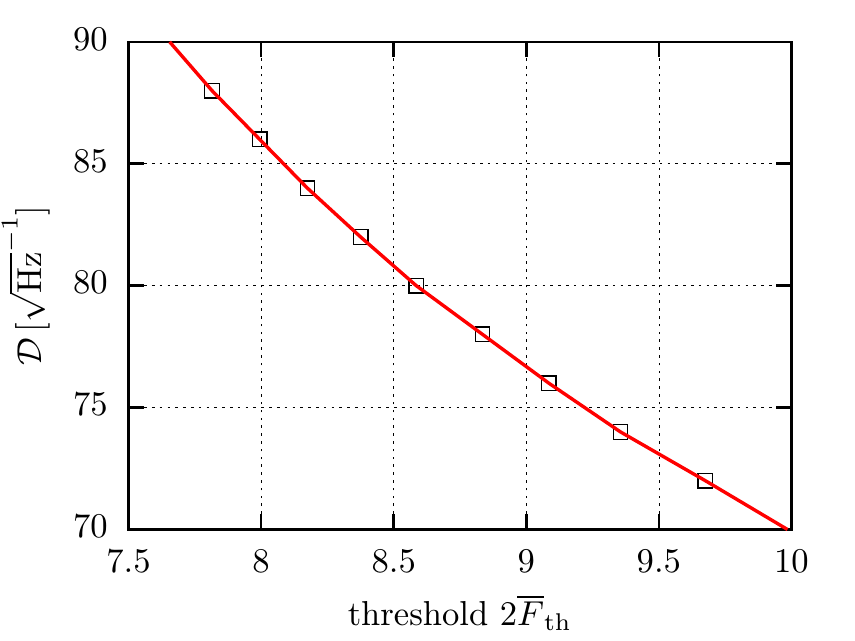}
  \caption{Sensitivity depth versus detection threshold. Boxes and solid lines indicate the piecewise-linear
    interpolation through the obtained thresholds at different depths of an injection-recovery simulation, using the
    \search{S6}{CasA}{\GCT} search setup (\cite{S6CasA} and Sec.~\ref{ssec: directed}).}
  \label{fig: linear}
\end{figure}

An alternative approach is used in this work to obtain the simulated ULs: for each set of fixed-$\depth$ injections and
recoveries, we determine the threshold on the statistic required in order to obtain the desired detection fraction $\pDET$.
This is illustrated in Fig.~\ref{fig: FindThreshold}, which shows a histogram of the observed loudest $2\avgFstat$
candidates obtained in each of $N=10^4$ injection and recovery runs at a fixed signal depth of $\depth=\SI{86}{\perRootHz}$,
using the \search{S6}{CasA}{\GCT} search setup (cf.\ Sec.~\ref{ssec: directed}).
By integrating the probability density from $2\avgFstat=0$ until we reach the desired value $1-\pDET$, we find the
detection threshold $2\avgFstat\thr$ at this signal depth $\depth$.
Repeating this procedure at different depths therefore generates a sampling of the function $\depth^{\pDET}(2\avgFstat\thr)$,
illustrated in Fig.~\ref{fig: linear}. These points can be interpolated to the required detection threshold, which
yields the desired upper-limit depth $\depth^{\pDET}$.

We see in in Fig.~\ref{fig: linear} that this function appears to be less ``curvy'' in the region of interest compared
to $\pDET(\depth)$ shown in Fig.~\ref{fig: Sigmoid}.
This allows for easier fitting and interpolation, for example a linear or quadratic fit should work quite well.
In fact, here we have simply used piecewise-linear interpolation, which is sufficient given our relatively fine
sampling of signal depths.

As already mentioned in the previous section, using the sensitivity estimate of
Sec.~\ref{sec:alt-notat-sens} one can determine the most relevant region of interest beforehand and focus the
Monte-Carlo injection-recoveries on this region, which will help ensure that any simple interpolation method will work
well.

Alternatively, for either the $\pDET(\depth)$- or the $\depth(2\FF\thr)$-sampling approach, one could also use an iterative
root-finding method to approach the desired $\pDET$ or $2\FF\thr$, respectively.

\section{\label{sec:accuracy} Comparing estimates against simulated upper limits}

In this section we compare the sensitivity estimates from Sec.~\ref{sec:alt-notat-sens} against simulated ULs for two
example cases (an all-sky search and a directed search), in order to quantify the accuracy and reliability of the
estimation method and implementation. This comparison shows generally good agreement, and also some instructive deviations.

Both examples are wide-parameter-space searches using a template bank over the unknown signal parameter dimensions
(namely, \{sky, frequency and spindown\} in the all-sky case, and \{frequency and first and second derivatives\} in the
directed-search case).

The simulated-UL procedure (see Sec.~\ref{sec:upperLimits}) performs a template-bank search over a box in parameter
space containing the injected signal (at a randomized location) in Gaussian noise.
On the other hand, the sensitivity estimate (cf.\ Eq.~\eqref{eq:19}) uses the mismatch distribution $\prior{\mis}$
obtained for this template bank via injection-recovery box searches on signals without noise.
We refer to this in the following as the \emph{box search}.

It will be instructive to also consider the (unrealistic) case of a perfectly-matched search, using only a single
template that matches the signal parameters perfectly for every injection, corresponding to zero mismatch $\mis=0$ in
Eq.~\eqref{eq:19}. We refer to this as the \emph{zero-mismatch search}.

\subsection{Example: \search{S6}{\AllSky}{\GCT} search}
\label{sec:all-sky-search}

In this example we use the setup of the all-sky search \search{S6}{\AllSky}{\GCT} \cite{S6Bucket}, which was using
the GCT implementation \cite{2009PhRvL.103r1102P} of the \StackSlideF{} statistic and was performed on the
volunteer-computing project Einstein@Home \cite{EinsteinAtHome}, see Table~\ref{tbl: allsky} and Sec.~\ref{ssec: allsky}
for more details.

\begin{figure}[htb]
  \centering
  \includegraphics[width = \columnwidth]{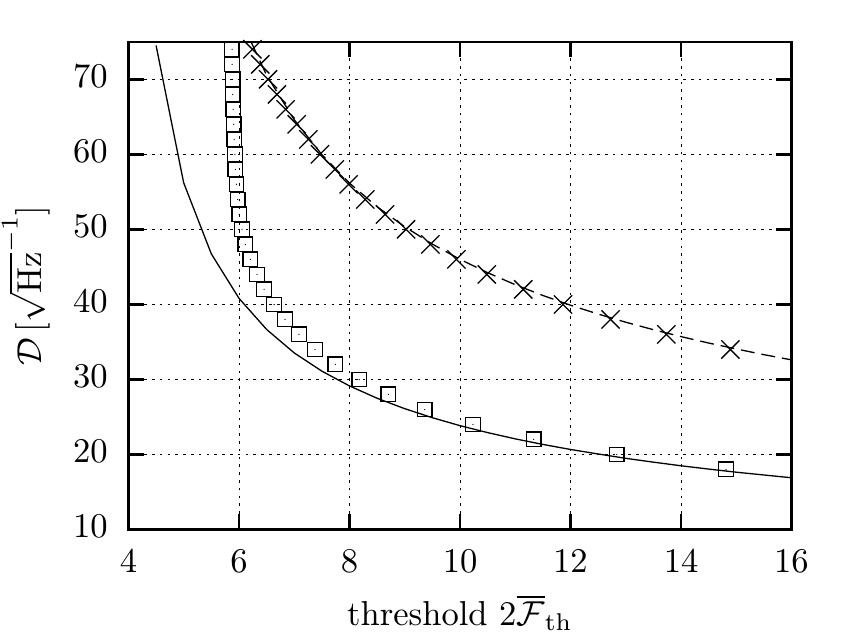}
  \caption{Comparison of estimated and simulated sensitivity depth $\depth^{90\%}$ as a function of threshold
    $2\avgFstat\thr$ for the \search{S6}{\AllSky}{\GCT} search \cite{S6Bucket}.
    The solid line shows the UL estimate for the \emph{box search}, and the squares ($\square$) show the
    corresponding simulated ULs.
    The dashed line indicates the estimate for the \emph{zero-mismatch case}, and the crosses ($\times$) are for the
    simulated zero-mismatch ULs. In the box search we observe an increasing divergence at decreasing thresholds due to
    noise effects, discussed in Sec.~\ref{sec:all-sky-search}.
  }
  \label{fig: EstMeasS6Bucket}
\end{figure}

Figure~\ref{fig: EstMeasS6Bucket} shows the comparison between simulated ULs and estimated sensitivity depths
$\depth^{90\%}$ versus threshold $2\avgFstat\thr$, for the \emph{box search} (squares and solid line), as well as for
the \emph{zero-mismatch search} (crosses and dashed line).
We see excellent agreement between estimated and simulated ULs for the zero-mismatch search. We also find very good
agreement for the box-search at higher thresholds, while we see an increasing divergence $\depth\rightarrow\infty$ of
the simulated ULs at decreasing thresholds, which is not captured by the estimate.

This discrepancy can be understood as the effect of noise fluctuations, which can enter in two different ways (that are
not completely independent of each other):
\begin{enumerate}[(i)]

\item For decreasing thresholds the corresponding false-alarm level Eq.~\eqref{eq:8a} grows, as it becomes increasingly
  likely that a ``pure noise'' candidate (i.e.\ unrelated to a signal) crosses the threshold.
In the extreme case where $\pFA$ approaches $\pDET$, the frequentist upper limit would tend to
$h_0\rightarrow 0$, corresponding to $\depth\rightarrow\infty$\footnote{Bayesian upper limits do not have this
  property, e.g.\ see \cite{FreqVsBayesUL} for more detailed analysis of these different types of upper limits.}.
This is illustrated in Fig.~\ref{fig: S6BucketNoise2F} showing the distribution of the loudest $2\avgFstat$ in a
box search on pure Gaussian noise, which can be compared to the diverging depth of the simulated box search around
$2\avgFstat\thr\lesssim6$ in Fig.~\ref{fig: EstMeasS6Bucket}.
\begin{figure}[htb]
  \includegraphics[width = \columnwidth]{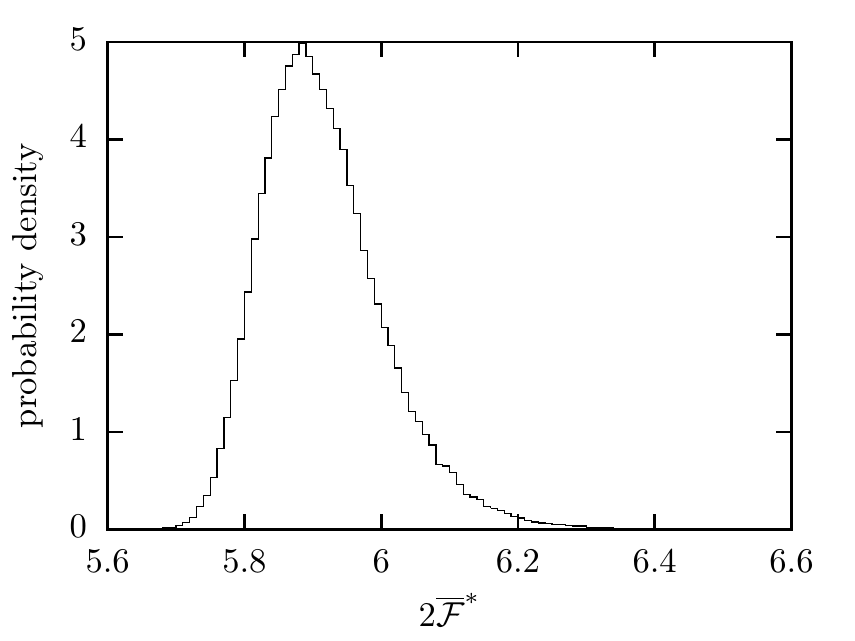}
  \caption{Distribution of the loudest $2\avgFstat$ for a box search on pure Gaussian noise, using the
    \search{S6}{\AllSky}{\GCT} search setup.
  }
  \label{fig: S6BucketNoise2F}
\end{figure}

We note that in practice the procedures used for \emph{measured ULs} in CW searches typically make sure that the
detection threshold has a very small false-alarm level, and we therefore expect this effect to have a negligible
impact in cases of practical interest.

\item The sensitivity estimate for wide-parameter-space searches makes the assumption that the loudest candidate
  $2\avgFstat\loudest$ is always found in the \emph{closest} template to the signal (i.e.\ with the smallest mismatch
  $\mis$), as discussed in Sec.~\ref{sec:appr-wide-param}.
  However, while the closest template has the highest \emph{expected} statistic value (by definition), other templates
  can actually produce the loudest statistic value in any given noise realization. How likely that is to happen depends
  on the details of the parameter space, the template bank and the threshold. It will typically be more likely at lower
  thresholds, as more templates further away from the signal are given a chance to cross the threshold
  (despite their larger mismatch).

  The \emph{true} distribution $\prob{2\avgFstat\loudest}{h_0}$ of a box search will therefore be shifted to higher
  values compared to the approximate distribution used in Eq.~\eqref{eq:17}. This implies that an actual search can have
  a higher detection probability than predicted by the estimate (corresponding to a larger sensitivity depth).
\end{enumerate}
Both of these effects contribute to different extents to the box-search discrepancy in Fig.~\ref{fig: EstMeasS6Bucket}
at lower thresholds:

The sampling distribution for $2\avgFstat\loudest$ in the presence of relatively strong signals at
$\depth=\SI{20}{\perRootHz}$ is shown in Fig.~\ref{fig: S6Bucket2FdistD20}, both for a simulated box search as well for
the assumed distribution in the estimate.
\begin{figure}[htb]
  \centering
  \includegraphics[width = \columnwidth]{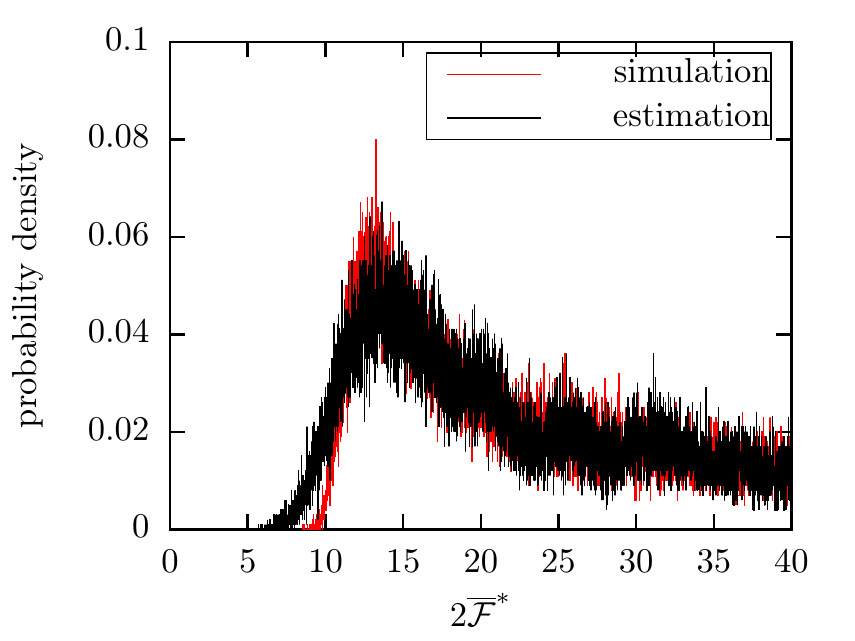}
  \caption{Loudest $2\avgFstat$ distribution for a box-search (using the \search{S6}{\AllSky}{\GCT} setup) with signals
    at a depth of $\depth=\SI{20}{\perRootHz}$.
    The black histogram shows the assumed distribution for sensitivity \emph{estimation} in Eq.~(\ref{eq:17}), and the
    lighter color shows the histogram obtained in a Monte-Carlo \emph{simulation} with signals injected in Gaussian noise.
  }
  \label{fig: S6Bucket2FdistD20}
\end{figure}
We see that most of the loudest candidates obtained in the simulation are above $2\avgFstat\loudest > 9$, and are
therefore extremely unlikely to be due to noise alone, as seen from Fig.~\ref{fig: S6BucketNoise2F}.
The difference between the two distributions in Fig.~\ref{fig: S6Bucket2FdistD20} is therefore soley due to effect (ii).
However, we see in Fig.~\ref{fig: EstMeasS6Bucket} that the resulting discrepancy in the sensitivity estimate at
$\depth=\SI{20}{\perRootHz}$ is still very small.

\begin{figure}[htb]
  \centering
  \includegraphics[width = \columnwidth]{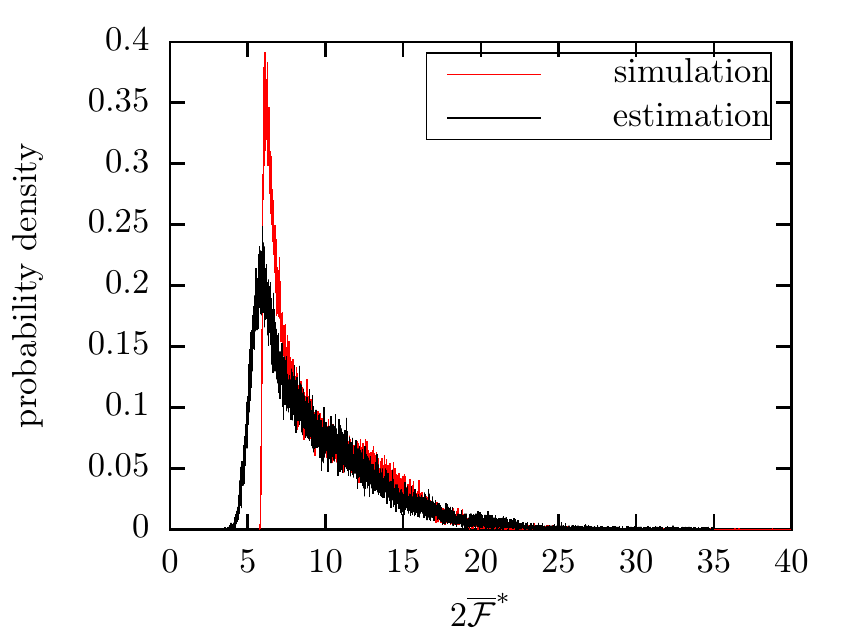}
  \caption{Same as Fig.~\ref{fig: S6Bucket2FdistD20} for weaker injected signals at a depth of $\depth=\SI{46}{\perRootHz}$.
  }
  \label{fig: S6Bucket2FdistD46}
\end{figure}

For weaker signals at $\depth=\SI{46}{\perRootHz}$, we see in Fig.~\ref{fig: S6Bucket2FdistD46} that the corresponding
distribution now overlaps with the pure-noise distribution of Fig.~\ref{fig: S6BucketNoise2F}.
The sensitivity depth therefore increasingly diverges for thresholds in the range $2\avgFstat\thr \sim [5.8, 6.1]$ due
to the increasing impact of effect (i).

\subsection{Example: multi-directed \search{O1}{MD}{\GCT}}
\label{sec:direct-search-exampl}

In this example we use the search setup of the directed search \search{O1}{MD}{\GCT} \cite{2018PhRvD..97b4051M}
currently running on Einstein@Home. This search consists of several directed searches for different targets on the sky,
including Vela~Jr.\ and Cas-A.

\begin{figure}[htb]
  \centering
  \includegraphics[width=\columnwidth]{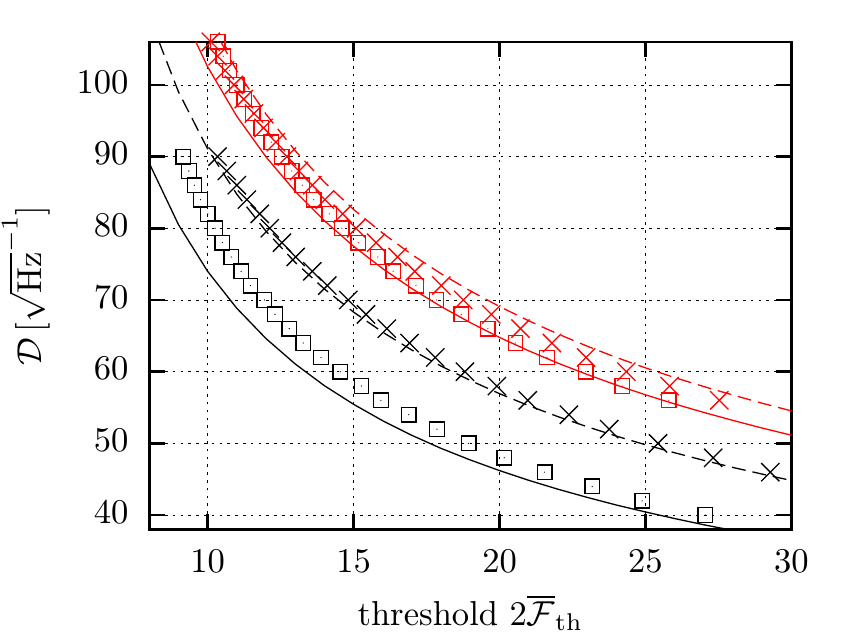}
  \caption{
    Comparison of estimated and simulated sensitivity depth $\depth^{90\%}$ as a function of the threshold
    $2\avgFstat\thr$ for two targets of the multi-directed search setup \search{O1}{MD}{\GCT}.
    The solid lines show the UL estimate for a \emph{box search}, while the squares ($\square$) show the
    corresponding simulated ULs.
    The dashed lines indicate the estimate for the \emph{zero-mismatch case}, and the crosses ($\times$) are for the
    simulated zero-mismatch ULs.
    The upper group of curves are for the target Vela~Jr., while the lower group of curves are for Cas~A.
  }
\label{fig: EstMeasO1MD}
\end{figure}

The comparison between simulated and estimated UL depths $\depth^{90\%}$ for these two targets is shown in
Fig.~\ref{fig: EstMeasO1MD}.
We see again very good agreement (relative deviations $\lesssim 3\%$) in the zero-mismatch case.
However, these deviations are larger than in the all-sky case shown in Fig.~\ref{fig: EstMeasS6Bucket}.
We suspect that this is due to the different antenna-pattern implementations of Eq.~\eqref{eq:51} between the search
code and the estimation scripts: we see different signs of the deviation for different sky positions (Vela~Jr.\ versus
Cas-A), and the effect disappears when averaging over the whole sky (as seen in Fig.~\ref{fig: EstMeasS6Bucket}).
However, the small size of the deviations did not warrant further efforts to try to mitigate this.

For the box-search case we see good agreement at higher thresholds, with again increasing deviations at lower thresholds
due to the noise effects discussed in the previous all-sky example Sec.~\ref{sec:all-sky-search}.

\section{\label{ssec: CompEstSim}Comparing estimates against measured upper limits}

In this section we present a general overview of measured sensitivity depths $\depth\meas$ derived from the published
upper limits of various past CW searches.
For the subset of searches where an $\FF$-statistic-based method was used (and for Bayesian targeted ULs), we provide
the sensitivity estimate for comparison.

The results are summarized in Tables \ref{tbl: allsky}--~\ref{tbl: binary} for the different search categories
(all-sky, directed and narrow-band, binary and targeted), and more details about each search are found in
Appendix~\ref{sec:SearchOverview}.

\subsection{General remarks and caveats}
\label{sec:gener-remarks-cave}

\subsubsection{Bayesian UL comparison}
\label{sec:bayes-ul-comp}

We also provide sensitivity estimates (using the framework of Sec.~\ref{sec:bayes-sens-estim}) for comparison to the
Bayesian ULs of targeted searches for known pulsars (\BayesPE), although these searches compute the $h_0$-posterior directly from
the data rather than from an $\FF$-statistic, which makes the comparison somewhat more indirect: we cannot use a known
threshold or loudest candidate $2\FF\loudest$, and we instead compute an \emph{expected depth} by calculating estimates
for $2\FF\loudest$-values drawn randomly from the central $\chi^2_4$-distribution and averaging the results.

\subsubsection{Converting published $h_0$ ULs into Depths $\depth$}
\label{sec:conv-publ-h_0}

Some searches already provide their upper limits in the form of a sensitivity depth $\depth^{\pDET}$, but in most cases
only the amplitude upper-limits $h_0^{\pDET}$ are given.
For these latter cases we try to use a reasonable PSD estimate $\Sn(f)$ for the data used in the search in order to
convert the quoted amplitude upper limits into sensitivity depths according to Eq.~\eqref{eq:10}.
This PSD estimate introduces a systematic uncertainty in the converted depth values, as in most cases we do not have
access to the ``original'' PSD estimate used for the $h_0$ UL calculation.

In particular, even small differences in windowing or the type of frequency averaging can results in large differences
in the PSD estimate near spectral disturbances. This can translate into large differences in the resulting converted
depth values. In order to mitigate outliers due to such noise artifacts we quote the \emph{median} over
the converted measured depth values $\{\depth_k\}$ (where $k$ either runs over multiple frequencies, targets or
detectors) and estimate the corresponding standard deviation using the \emph{mean absolute deviation} (MAD)
\cite{RousseeuwCroux93:_MAD}, namely
\begin{equation}
  \label{eq:15}
  \begin{split}
    \depthMED &\equiv \median\left[ \depth_k\right]\,,\\
    \sigmaMAD &\equiv 1.4826 \, \median\left[\, \left|\depth_k - \depthMED\right| \,\right]\,.
  \end{split}
\end{equation}

\subsubsection{Comparing different searches by sensitivity depth $\depth$}
\label{sec:comp-diff-search}

We can see in the tables \ref{tbl: allsky}--~\ref{tbl: binary} that searches within the same search category often show
roughly comparable sensitivity depths.
At one end of the spectrum are the fully-targeted searches, for which the parameter space (for each pulsar) is a single
point, and one can achieve the maximal possible sensitivity for the available data, namely
$\depth\sim\Ord{\SI{500}{\perRootHz}}$ (see Table~\ref{tbl: targeted}).
At the other end of the spectrum lies the all-sky binary search with a sensitivity depth of
$\depth\sim\SI{3}{\perRootHz}$ (see Table~\ref{tbl: binary}), which covers the largest parameter space of any search to
date.

One cannot directly compare searches on sensitivity depth alone, even within the same
search category. Other key aspects of a search are the parameter-space volume covered, the total computing power used,
and the robustness of the search to deviations from the assumed signal- or noise-model.

Is it intuitively obvious that the more computing power spent on a fixed parameter-space volume, the more sensitive the
search will tend to be, although the increase in sensitivity is typically very weak, often of order the 10th-14th root
of the computing power \cite{prix2012optimal}.

It is also evident that the larger the parameter space covered by a search, the less sensitivity depth can be achieved
due to the increased spending of computing power on ``breadth'' rather than depth.
Ultimately the most directly relevant characteristic of a search would be its \emph{total detection probability}
\cite{2016PhRvD..93f4011M,2018PhRvD..97b4051M}, which factors in both breadth and depth as well as the underlying
astrophysical prior on signal amplitudes over the parameter space searched.

\subsection{\label{ssec: CompAllSky}All-sky searches}

Estimated and measured sensitivity depths for all-sky searches are given in Table~\ref{tbl: allsky}, and further details
about individual searches can be found in appendix~\ref{ssec: allsky}.
\begin{table*}
  \caption{All-sky searches: estimated $\depth\est$ and measured sensitivity depth $\depth\meas$
    (median and standard deviation, see Sec.~\ref{sec:conv-publ-h_0}).
    The columns labeled $f$ and $\dot{f}$ give the frequency and spindown ranges covered by each search.
    Sensitivity depths in \textit{italics} refer to $90\%$-confidence upper limits, while normal font refers to
    $95\%$-confidence.
    See appendix~\ref{ssec: allsky} for further details on the individual results.
  }
  \begin{ruledtabular}
\begin{tabular}{ccccccccc} 
 Data & Search method &$f\,[\si{Hz}]$ & $\dot{f}\,[\si{\nano\hertz\per\second}]$ & $\depth\est\,[\si{\perRootHz}]$ & $\depthMED\meas\,[\si{\perRootHz}]$ & $\sigmaMAD_{\depth\meas}\,[\si{\perRootHz}]$ & Ref, Sec \\[1 mm] 
\hline \\[-1 mm] 
S2 & Hough & [200, 400] & [-1.1, 0] & -- & 11.3 & 1.5 & \cite{S2AS},\ref{sec:S2-AS-Hough} \\ 
\vspace{2 mm}S2 & \Fstat & [160, 728.8] & 0 & 6.5 & 5.5 & 1.6 & \cite{S2ScoX1+AS},\ref{sec:S2-AS-Fstat} \\ 
S4 & StackSlide & [50, 1000] & [-10, 0] & -- & 10.5 & 1.1 & \cite{S4AS},\ref{sec:S4-AS-MultiPipeline} \\ 
S4 & Hough & [50, 1000] & [-2.2, 0] & -- & 13.4 & 0.7 & \cite{S4AS},\ref{sec:S4-AS-MultiPipeline} \\ 
S4 & PowerFlux & [50, 1000] & [-10, 0] & -- & \{6.1, 21.3\}\footnotemark[1] & \{0.7, 2.3\} & \cite{S4AS},\ref{sec:S4-AS-MultiPipeline} \\ 
\vspace{2 mm}S4 & \Fstat\Coinc & [50, 1500] & [-9.5, 1] & -- & \ninety{8.5} & 0.5 & \cite{S4E@HAS},\ref{sec:S4-AS-FstatCoinc} \\ 
earlyS5 & PowerFlux & [50, 1100] & [-5, 0] & -- & \{16.1, 47.9\}\footnotemark[1] & \{2.4, 5.9\} & \cite{S5Early},\ref{sec:earlyS5-AS-PowerFlux} \\ 
\vspace{2 mm}earlyS5 & \Fstat\Coinc & [50, 1500] & [-12.7, 1.3] & -- & \ninety{10.9} & 0.2 & \cite{S5E@HEarly},\ref{sec:earlyS5-AS-FstatCoinc} \\ 
S5 & PowerFlux & [50, 800] & [-6, 0] & -- & \{25.7, 71.3\}\footnotemark[1] & \{0.7, 2.2\} & \cite{S5AS},\ref{sec:S5-AS-PowerFlux} \\ 
S5 & \HoughF & [50, 1190] & [-2, 0.1] & \ninety{30.5} & \ninety{30.0} & 1.4 & \cite{S5R5},\ref{sec:S5-AS-HoughF} \\ 
S5 & Hough & [50, 1000] & [-0.9, 0] & -- & 28.1 & 0.6 & \cite{S5Hough},\ref{sec:S5-AS-Hough} \\ 
\vspace{2 mm}S5 & \GCT & [1249.7, 1499.7] & [-2.9, 0.6] & \ninety{27.0} & \ninety{30.7} & -- & \cite{S5GC1HF},\ref{sec:S5-AS-EaH} \\ 
\vspace{2 mm}VSR1 & \TDFstat & [100, 1000] & [-16, 0] & -- & \ninety{22.6} & {6.0} & \cite{VSR1},\ref{sec:VSR1-AS-TDFstat} \\ 
\vspace{2 mm}VSR2,4 & FreqHough+FUP & [20, 128] & [-0.1, 0.015] & -- & \ninety{35.5} & 11.1 & \cite{VSR24LF},\ref{sec:VSR24-AS-FreqHough} \\ 
S6 & \GCT & [50, 510] & [-2.7, 0.3] & \ninety{34.4} & \ninety{37.0} & -- & \cite{S6Bucket},\ref{sec:S6-AS-EaH} \\ 
S6 & \GCT+\FollowUp & [50, 510] & [-2.7, 0.3] & \ninety{38.3} & \ninety{46.9} & -- & \cite{S6BucketFU},\ref{sec:S6-AS-EaH-FUP} \\ 
\vspace{2 mm}S6 & PowerFlux & [100, 1500] & [-11.8, 10] & -- & \{17.9, 52.8\}\footnotemark[1] & \{1.4, 3.4\} & \cite{S6CompPF},\ref{sec:S6-AS-PowerFlux} \\ 
O1 & \GCT & [20, 100] & [-2.7, 0.3] & \ninety{46.4} & \ninety{48.7} & -- & \cite{O1ASE@H},\ref{sec:O1-AS-EaH} \\ 
O1 & PowerFlux & [20, 200] & [-10, 1] & -- & \ninety{28.9} & 2.2 & \cite{O1ASE@H},\ref{sec:O1-AS-MultiPipeline} \\ 
O1 & PowerFlux & [20, 475] & [-10, 1] & -- & \{19.9, 54.6\}\footnotemark[1] & \{1.3, 3.2\} & \cite{O1AS},\ref{sec:O1-AS-MultiPipeline} \\ 
O1 & SkyHough & [20, 475] & [-10, 1] & -- & 22.4 & 1.1 & \cite{O1AS},\ref{sec:O1-AS-MultiPipeline} \\ 
O1 & \TDFstat & [20, 475] & [-10, 1] & -- & 23.7 & 2.1 & \cite{O1AS},\ref{sec:O1-AS-MultiPipeline} \\ 
O1 & FreqHough & [20, 475] & [-10, 1] & -- & 21.4 & 10.6 & \cite{O1AS},\ref{sec:O1-AS-MultiPipeline} \\ 
O1 & PowerFlux & [475, 2000] & [-10, 1] & -- & \{18.6, 50.9\}\footnotemark[1] & \{1.3, 3.4\} & \cite{O1ASFullBand},\ref{sec:O1-AS-MultiPipeline} \\ 
O1 & SkyHough & [475, 2000] & [-10, 1] & -- & 16.8 & 3.0 & \cite{O1ASFullBand},\ref{sec:O1-AS-MultiPipeline} \\ 
O1 & \TDFstat & [475, 2000] & [-10, 1] & -- & 10.9 & 0.6 & \cite{O1ASFullBand},\ref{sec:O1-AS-MultiPipeline} \\ 
\end{tabular} 
\footnotetext[1]{Sensitivity depths corresponding to worst linear and circular polarization, respectively, cf.~Sec.~\ref{ssec:DetailsIntro}}

    \label{tbl: allsky}
  \end{ruledtabular}
\end{table*}
\newcommand{\AllSkyMeanAcc}{9}
\newcommand{\AllSkyMedianAcc}{7}

The mean relative error between measured and estimated depths is $\SI{\AllSkyMeanAcc}{\%}$, while the median error is
$\SI{\AllSkyMedianAcc}{\%}$.

One case of interest is the surprisingly large discrepancy of $\sim 18\%$ observed for the
\search{S6}{\AllSky}{\GCT+\FollowUp} search, shown in Fig.~\ref{fig: EstMeasS6Bucket}, were we see a significantly
higher measured depth ($\depthMED\meas=\SI{46.9}{\perRootHz}$) than estimated ($\depth\est=\SI{38.3}{\perRootHz}$).
This can be traced back to the template-maximization approximation used in the estimate, namely effect (ii) discussed in
Sec.~\ref{sec:all-sky-search}.
The low threshold used in the search ($2\avgFstat\thr=6.1$) appears to be at the cusp of becoming affected by pure-noise
candidates (effect (i) in Sec.~\ref{sec:all-sky-search}), but this effect is still small and does not account for the
discrepancy. Furthermore, the upper limit procedure used a multi-stage follow-up, which ensures the final false-alarm
level (p-value) is very small, which rules out contamination from pure-noise candidates.
\begin{figure}[htb]
  \includegraphics[width = \columnwidth]{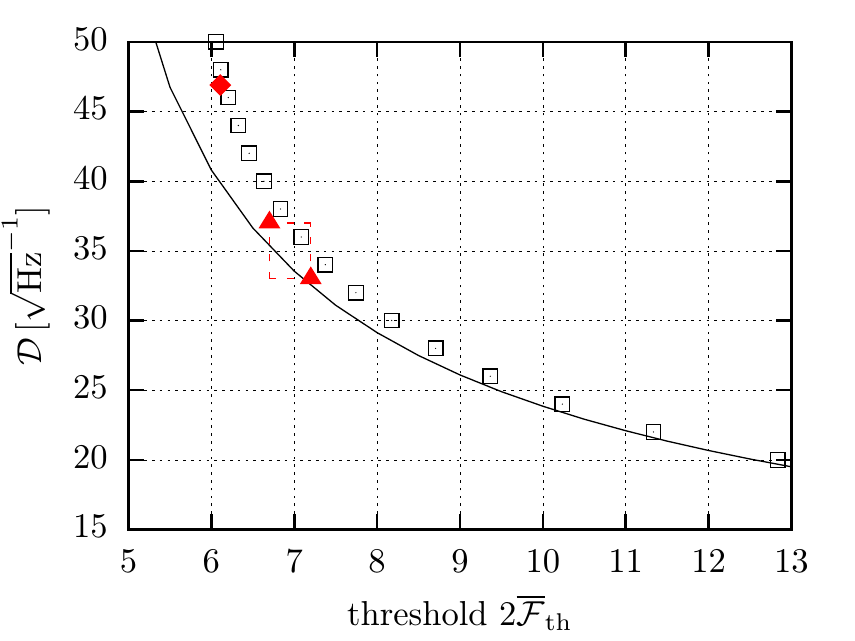}
  \caption{
    Estimated (--) and simulated ($\square$) sensitivity depth versus threshold $2\avgFstat\thr$ for the
    \search{S6}{\AllSky}{\GCT} (\method{+\FollowUp}) search setup, illustrating the effect of the template-maximization
    in the estimate (discussed in Sec.~\ref{sec:all-sky-search}).
    The triangles ($\Delta$) and dashed lines show the measured upper-limit depth $\depthMED\meas$ in the initial
    \search{S6}{\AllSky}{\GCT} search \cite{S6Bucket}, and the diamond ($\diamond$) shows the corresponding result from
    the follow-up (\method{\FollowUp}) search \cite{S6BucketFU} (threshold $2\avgFstat\thr=6.1$).
  }
  \label{fig: EstPaperS6Bucket}
\end{figure}

\subsection{\label{ssec: CompDirected}Directed and Narrow-band searches}

Estimated and measured sensitivity depths for directed and narrow-band searches are given in Tables~\ref{tbl: directed}
and \ref{tbl: nineyoung}, and further details about individual searches can be found in appendix~\ref{ssec: directed}.
\begin{table*}[htb]
  \caption{Directed and narrow-band searches: estimated $\depth\est$ and measured sensitivity depth $\depth\meas$
    (median and standard deviation, see Sec.~\ref{sec:conv-publ-h_0}).
    The column labeled $f$ gives the frequency range covered by each search (omitting $\dot{f}$ and $\ddot{f}$ search ranges).
    Sensitivity depths in \textit{italics} refer to $90\%$-confidence upper limits, while normal font refers to
    $95\%$-confidence.
    See appendix~\ref{ssec: directed} for further details on the individual results.
  }
  \begin{ruledtabular}
\begin{tabular}{ccccccccc} 
 Science run & Search method & Target & $f\,[\si{Hz}]$ & $\depth\est\,[\si{\perRootHz}]$ & $\depthMED\meas\,[\si{\perRootHz}]$ & $\sigmaMAD_{\depth\meas}\,[\si{\perRootHz}]$ & Ref, Sec \\[1 mm] 
\hline \\[-1 mm] 
\vspace{2 mm}earlyS5 & \Fstat & Crab & 59.56$\pm$0.006 & 221.3 & 223.1 & -- & \cite{S5Crab},\ref{sec:S5-Crab-Fstat}\\ 
S5 & \Fstat & CasA & [100, 300] & 35.9 & 35.5 & 0.8 & \cite{S5CasA},\ref{sec:S5-CasA-Fstat}\\ 
\vspace{2 mm}S5 & \GCT & GalacticCenter & [78, 496] & \ninety{58.2} & \ninety{72.1} & 4.5 & \cite{S5GalacticCenter},\ref{sec:S5-GC-GCT}\\ 
VSR4 & 5-vector & Vela & 22.384$\pm$0.02 & -- & 100.5 & -- & \cite{VSR4VelaCrab},\ref{sec:VSR4-VelaCrab-5vector}\\ 
\vspace{2 mm}VSR4 & 5-vector & Crab & 59.445$\pm$0.02 & -- & 90.1 & -- & \cite{VSR4VelaCrab},\ref{sec:VSR4-VelaCrab-5vector}\\ 
S6 & \Fstat & NineYoung (table~\ref{tbl: nineyoung}) & [46, 2034] & 37.8 & 37.7 & 0.3 & \cite{S6NineYoung},\ref{sec:S6-NineYoung-Fstat}\\ 
S6 & \GCT & CasA & [50, 1000] & \ninety{79.6} & \ninety{72.9} & 0.4 & \cite{S6CasA},\ref{sec:S6-CasA-EaH}\\ 
S6 & LooselyCoherent & OrionSpur & [50, 1500] & -- & \{30.2, 85.7\}\footnotemark[2] & \{2.3, 4.3\} & \cite{S6OrionSpur},\ref{sec:S6-OrionSpur-PowerFlux}\\ 
\vspace{2 mm}S6 & \Fstat & NGC6544 & [92.5, 675] & 29.3 & 29.6 & 1.7 & \cite{S6GlobClust},\ref{sec:S6-NGC6544-Fstat}\\ 
O1 & 5-vector & 11 pulsars & $< \pm 0.1$\footnote[1]{search band around twice the pulsar spin frequency} & -- & 111.6 & 12.2 & \cite{O1Narrowband},\ref{sec:O1-NarrowBand-5vector}\\ 
O1 & Radiometer & SN1987A & [25, 1726] & -- & \ninety{11.1} & 4.3 & \cite{O1RadiometerDirected},\ref{sec:O1-Radiometer}\\ 
O1 & Radiometer & GalacticCenter & [25, 1726] & -- & \ninety{7.7} & 2.9 & \cite{O1RadiometerDirected},\ref{sec:O1-Radiometer}\\ 
\end{tabular} 
\footnotetext[2]{Sensitivity depths corresponding to worst linear and circular polarization, respectively, cf.~Sec.~\ref{ssec:DetailsIntro}}

    \label{tbl: directed}
  \end{ruledtabular}
\end{table*}
\newcommand{\DirectedMeanAcc}{5}
\newcommand{\DirectedMedianAcc}{1}

The mean relative error between measured and estimated depths is $\SI{\DirectedMeanAcc}{\%}$, and the median error is $\SI{\DirectedMedianAcc}{\%}$.

\begin{table*}[htb]
  \centering
  \caption{\search{S6}{NineYoung}{\Fstat} search: estimated $\depth\est$ and measured sensitivity depth $\depth\meas$
    (median and standard deviation, see Sec.~\ref{sec:conv-publ-h_0}).
    for nine young supernova remnants \cite{S6NineYoung}.
    All sensitivity depths refer to $95\%$-confidence.
    See appendix~\ref{sec:S6-NineYoung-Fstat} for further details.
  }
  \begin{ruledtabular}
\begin{tabular}{l |cccccccccc} 
SN remnant & G1.9 & G18.9 & G93.3 & G111.7 & G189.1 & G266.2deep & G266.2wide & G291.0 & G347.3 & G350.1\\ 
Name &  &  & DA 530 & Cas A & IC 443 & Vela Jr. & Vela Jr. & MSH 11-62 &  & \\ \hline\noalign{\smallskip} 
$\depth\est\,[\si{\perRootHz}]$ & $29.0$ & $43.9$ & $46.8$ & $29.3$ & $40.1$ & $38.3$ & $24.2$ & $41.1$ & $32.8$ & $37.3$\\[1mm] 
$\depthMED\meas\,[\si{\perRootHz}]$  & $28.3$ & $44.4$ & $49.6$ & $31.5$ & $39.2$ & $40.8$ & $26.1$ & $44.0$ & $32.1$ & $36.1$\\[1mm] 
$\sigmaMAD_{\depth\meas}\,[\si{\perRootHz}]$  & $0.8$ & $1.3$ & $1.5$ & $0.9$ & $1.2$ & $1.0$ & $0.7$ & $1.2$ & $0.8$ & $1.1$\\[3mm] 
$\Tdata\,[\SI{e6}{s}]$ & $1.2$ & $3.1$ & $2.8$ & $1.1$ & $2.3$ & $1.9$ & $0.7$ & $2.2$ & $1.4$ & $1.9$\\ 
$2\Fth$  & $ 58.0$ & $ 56.3$ & $ 55.6$ & $ 55.6$ & $ 55.3$ & $ 53.7$ & $ 52.8$ & $ 56.6$ & $ 54.1$ & $ 57.6$\\ 
\end{tabular} 

    \label{tbl: nineyoung}
  \end{ruledtabular}
\end{table*}
\newcommand{\NineYoungMeanAcc}{4}
\newcommand{\NineYoungMedianAcc}{4}

For the \search{S6}{NineYoung}{\Fstat} search for nine young supernova remnants shown in Table~\ref{tbl: nineyoung}, the mean relative
error between measured and estimated depths is $\SI{\NineYoungMeanAcc}{\%}$ (median error
$\SI{\NineYoungMedianAcc}{\%}$).

For two cases of interest we investigated more closely to understand the origin of the observed deviation:


\search{S5}{GalacticCenter}{\GCT} search \cite{S5GalacticCenter}:
the reason for the relatively large deviation of $19\%$ in this case between $\depth\est=\SI{58.2}{\perRootHz}$ and
$\depthMED\meas=\SI{72.1}{\perRootHz}$ can be understood by looking at the details of this search setup: contrary to the
assumed uniform averaging of antenna-pattern functions over time (cf.\ Sec.~\ref{sec:single-stage-stacksl}, this search setup was
specifically optimized by choosing the relatively short segments of $\Tseg=\SI{11.5}{hours}$ in such a way as to
maximize sensitivity, by selecting times of maximal antenna-pattern sensitivity towards the particular sky direction of the galactic
center. This is described in more detail in \cite{behnke2015postprocessing}, and is quoted there as yielding a sensitivity
improvement of about $20\%$, consistent with the observed enhancement of measured sensitivity compared to our
estimate.

\begin{figure}[htbp]
  \centering
  \includegraphics[width = \columnwidth]{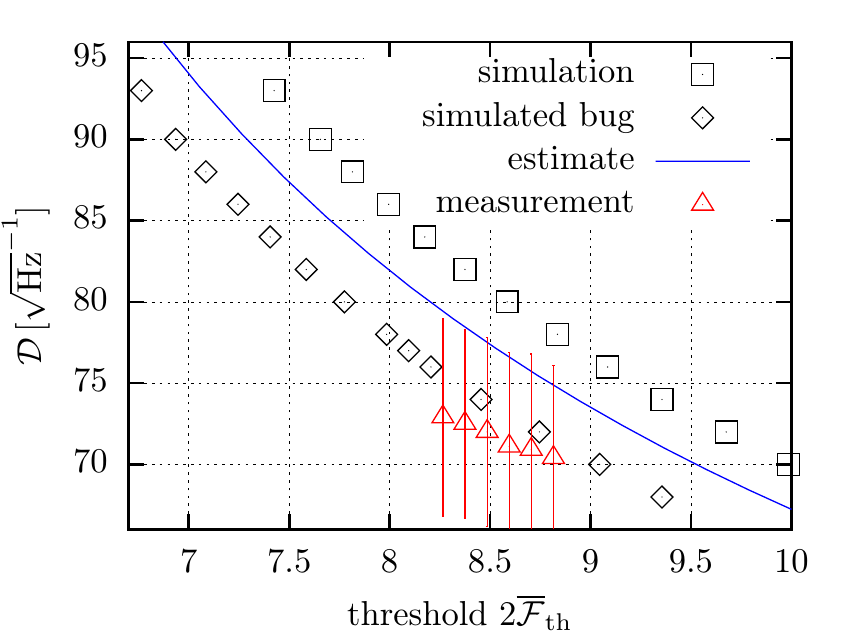}
  \caption{
    Estimated (--) and simulated ($\square$) sensitivity depth versus threshold $2\avgFstat\thr$ for the \search{S6}{CasA}{\GCT}
    search setup \cite{S6CasA}.
    The published upper limits are plotted as triangles ($\Delta$), while the diamonds ($\diamond$) show the
    simulated depths if we incorporate the bug found in the original UL procedure.
  }
  \label{fig: EstPaperS6CasA}
\end{figure}

\search{S6}{CasA}{\GCT} search \cite{S6CasA}:
the deviation between $\depth\est=\SI{79.6}{\perRootHz}$ versus $\depthMED\meas=\SI{72.9}{\perRootHz}$ does not
seem very large per se, but is unusual for the estimate typically does not tend to \emph{overestimate} sensitivity by
that much.
A detailed investigation led us to discover a bug in the original upper-limit script used in \cite{S6CasA}, which
resulted in the injection-recovery procedure to sometimes search the wrong box in parameter space, missing the
injected signal. By artificially reproducing the bug in our upper limit simulation we are able to confirm that this bug
does account for a decrease in detection probability of about $7\%$, resulting in an underestimate of the upper-limit
depth as shown in Fig.~\ref{fig: EstPaperS6CasA}.

\subsection{\label{ssec: CompBinary} Searches for neutron stars in binaries}

\newcommand{\BinaryMeanAcc}{8}
\newcommand{\BinaryMedianAcc}{8}
\newcommand{\sensDepthStwoFstatSco}{4.4}
Estimated and measured sensitivity depths for searches for CWs from neutron stars in binary systems are given in
Tables~\ref{tbl: binary}, and further details about individual searches can be found in appendix~\ref{ssec: binary}.
In this case the only $\FF$-statistic-based search is \search{S2}{ScoX1}{\Fstat}, for which we obtain an estimate of
$\depth\est = \SI{\sensDepthStwoFstatSco}{\perRootHz}$ (assuming an average mismatch of $\mu\sim0.1/3$ corresponding to
a cubic lattice with maximal mismatch of $0.1$ \cite{S2ScoX1+AS}). The relative error is between measured and estimated
sensitivity depth is therefore $\SI{\BinaryMeanAcc}{\%}$.

\begin{table*}
  \caption{Binary searches: measured sensitivity depth $\depth\meas$
    (median and standard deviation, see Sec.~\ref{sec:conv-publ-h_0}).
    All sensitivity depths refer to $95\%$-confidence.
    See appendix~\ref{ssec: binary} for further details on the individual results.
  }
  \begin{ruledtabular}
\begin{tabular}{ccccccc} 
 Science run & Search method & Target & f\,[\si{Hz}]& $\depthMED\meas\,[\si{\perRootHz}]$ & $\sigmaMAD_{\depth\meas}\,[\si{\perRootHz}]$ & Ref, Sec \\[1 mm] 
\hline \\[-1 mm] 
\vspace{2 mm}S2 & \Fstat & ScoX1 & [464, 484],[604, 624] & 4.1 & 0.1 & \cite{S2ScoX1+AS},\ref{sec:S2-ScoX1-Fstat} \\ 
\vspace{2 mm}S5 & Sideband & ScoX1 & [50, 550] & 8.1 & 1.0 & \cite{S5ScoX1},\ref{sec:S5-ScoX1-Sideband} \\ 
S6,VSR2,3 & TwoSpect & AllSky & [20, 520] & 3.2 & 0.4 & \cite{S6Binary},\ref{sec:S6VSR23-ASScoX1-TwoSpect} \\ 
S6,VSR2,3 & TwoSpect & ScoX1 & [20, 57.25] & 8.2 & 4.0 & \cite{S6Binary},\ref{sec:S6VSR23-ASScoX1-TwoSpect} \\ 
S6 & TwoSpect & ScoX1 & [40, 2040] & 5.7 & 1.6 & \cite{S6ScoX1},\ref{sec:S6-ScoX1-TwoSpectGrant} \\ 
\vspace{2 mm}S6 & TwoSpect & J1751 & \{435.5, 621.5, 870.5\}$\pm1$ & 9.4 & 1.2 & \cite{S6ScoX1},\ref{sec:S6-ScoX1-TwoSpectGrant} \\ 
O1 & Viterbi & ScoX1 & [60, 650] & 7.6 & 1.0 & \cite{O1ScoX1},\ref{sec:O1-ScoX1-Viterbi} \\ 
O1 & CrossCorr & ScoX1 & [25, 2000] & 24.0 & 2.0 & \cite{O1ScoX1CrossCorr},\ref{sec:O1-ScoX1-CrossCorr} \\ 
O1 & Radiometer & ScoX1 & [25, 1726] & 5.8 & 1.0 & \cite{O1RadiometerDirected},\ref{sec:O1-Radiometer} \\ 
\end{tabular} 

    \label{tbl: binary}
  \end{ruledtabular}
\end{table*}

\subsection{\label{ssec: CompTargeted} Targeted searches for known pulsars}
\newcommand{\depthVelaAnglePriors}{462.1\pm35.0}
\newcommand{\TargetedMeanAcc}{16}
\newcommand{\TargetedMedianAcc}{10}

Estimated and measured sensitivity depths for targeted searches are given in Tables~\ref{tbl: targeted}, and further
details about individual searches can be found in appendix~\ref{ssec: targeted}.

Note that the quoted upper limits of the \BayesPE{}-method are obtained by Bayesian parameter-estimation
\cite{BayesianPE} of $\prob{h_0}{x}$ directly on the data $x$.
Therefore we cannot directly apply the Bayesian sensitivity estimate derived in Sec.~\ref{sec:bayes-sens-estim},
which assumes an initial $\FF(x)$-statistic computed on the data, from which the Bayesian upper limit would be derived.
We therefore provide an approximate comparison with the \emph{expected} sensitivity estimate, which we compute by
estimating depths using $2\FF\loudest$ drawn from a central $\chi^2_4$ distribution (given each target
corresponds to a single template) and averaging the resulting estimated $\depth$ values.
In cases where several targets are covered by the search, we assume for simplicity that the targets are isotropically
distributed over the sky and compute a single all-sky sensitivity estimate. 
For single-target searches the exact sky position is used for the estimate.
\begin{table*}[htbp]
  \caption{Targeted searches for known pulsars: estimated $\depth\est$
    and measured sensitivity depth $\depth\meas$
    (with respectively, median and standard deviation, see Sec.~\ref{sec:bayes-ul-comp},\ref{sec:conv-publ-h_0}).
    All sensitivity depths refer to $95\%$-confidence.
    See appendix~\ref{ssec: targeted} for further details on the individual results.
  }
  \begin{ruledtabular}
\begin{tabular}{ccccccccc} 
 Science run & Search method & Targets & $\depthMED\est\,[\si{\perRootHz}]$ & $\sigmaMAD_{\depth\est}\, [\si{\perRootHz}]$ & $\depthMED\meas\,[\si{\perRootHz}]$ & $\sigmaMAD_{\depth\meas}\,[\si{\perRootHz}]$ & Ref, Sec \\[1 mm] 
\hline \\[-1 mm] 
S1 & \Fstat (worst-orientation) & J1939+21 & 70.8 & 39.8 & 64.2 & 38.1 & \cite{S1J1939},\ref{sec:S1-J1939-FstatBayesPE} \\ 
S1 & \Fstat & J1939+21 & 110.4 & 66.7 & 101.8 & 61.8 & \cite{S1J1939},\ref{sec:S1-J1939-FstatBayesPE} \\ 
\vspace{2 mm}S1 & \BayesPE & J1939+21 & 81.5 & 19.8 & 85.2 & 14.3 & \cite{S1J1939},\ref{sec:S1-J1939-FstatBayesPE} \\ 
\vspace{2 mm}S2 & \BayesPE & 28 pulsars & 243.5 & 54.3 & 156.4 & 42.2 & \cite{S2KnownPulsar},\ref{sec:S2-KnownPulsars-BayesPE} \\ 
\vspace{2 mm}S3,4 & \BayesPE & 78 pulsars & 337.8 & 81.2 & 299.5 & 79.0 & \cite{S3S4KnownPulsar},\ref{sec:S34-KnownPulsars-BayesPE} \\ 
earlyS5 & \BayesPE & Crab & 621.3 & 129.7 & 774.1 & -- & \cite{S5Crab},\ref{sec:earlyS5-Crab-BayesPE} \\ 
\vspace{2 mm}S5 & \BayesPE & 116 pulsars & 997.8 & 210.4 & 932.1 & 317.1 & \cite{S5KnownPulsar},\ref{sec:S5-KnownPulsars-BayesPE} \\ 
\vspace{2 mm}VSR2 & \BayesPE,\Fstat,5-vector & Vela & 351.9 & 78.5 & 408.5 & 20.8 & \cite{VSR2Vela},\ref{sec:VSR2-Vela-MultiMethod} \\ 
\vspace{2 mm}S6,VSR2,4 & \BayesPE,\Fstat,5-vector & 195 pulsars & 555.7 & 116.2 & 514.7 & 171.0 & \cite{S6KnownPulsar},\ref{sec:S6VSR24-KnownPulsar-MultiMethod} \\ 
O1 & \BayesPE,\Fstat,5-vector & 200 pulsars & 321.6 & 74.0 & 355.8 & 95.4 & \cite{O1KnownPulsar},\ref{sec:O1-KnownPulsar-MultiMethod} \\ 
\end{tabular} 

    \label{tbl: targeted}
  \end{ruledtabular}
\end{table*}
The mean relative error between measured and estimated depths is $\SI{\TargetedMeanAcc}{\%}$, and the median error is $\SI{\TargetedMedianAcc}{\%}$.

\section{Discussion}
\label{sec:discussion}

In this paper we presented a fast and accurate sensitivity-estimation framework and implementation for
$\FF$-statistic-based search methods for continuous gravitational waves, extending and generalizing an earlier analytic
estimate derived by \citet{wette2012estimating}.
In particular the new method is more direct and uses fewer approximations for single-stage \StackSlideF{} searches, and
is also applicable to multi-stage \StackSlideF{} searches, \HoughF{} searches and Bayesian upper limits (based on
$\FF$-statistic searches).

The typical runtime per sensitivity estimate is about $\SI{10}{seconds}$ with cached $\prior{R^2}$ distribution, and
about \SI{25}{seconds} per detector for the first call with a new parameter prior.
The accuracy compared to simulated Monte-Carlo upper limits in Gaussian noise is within a few $\%$ (provided the
threshold corresponds to a low false-alarm level), and we find generally good agreement (of less than $\sim 10\%$ average
error) compared to published upper limits in the literature.
Several factors leading to the observed deviations in various cases are discussed in detail.

We also provided a comprehensive overview of published CW upper limit results, converting the quoted $h_0$ upper limits
into sensitivity depths. This introduces some systematic uncertainties, as we often do not have access to the original
PSD estimate used for the upper limits. We therefore advocate for future searches to directly provide their upper-limit
results also in terms of the sensitivity depth of Eq.~\eqref{eq:10}, in order to allow easier direct comparison between
searches and to sensitivity estimates.

\begin{acknowledgments}
We thank Sylvia Zhu and Heinz-Bernd Eggenstein for help in recovering information for past Einstein@Home searches, and
Sylvia in particular for helping to localize the bug in the \search{S6}{CasA}{\GCT} upper limit.
We thank Vladimir Dergachev for help with the PowerFlux upper limits, Chris Messenger with helping us interpret the
SideBand upper limits, and and John T.\ Whelan, Sinéad Walsh and Avneet Singh for helpful comments.
Numerical calculations were performed on the ATLAS computing cluster of the Albert-Einstein Institute in Hannover.
KW is supported by Australian Research Council grant CE170100004.
This document has been assigned LIGO document number \dcc.
\end{acknowledgments}
\begin{appendix}
\section{\label{sec:SearchOverview} Details on referenced CW searches}
\subsection{\label{ssec:DetailsIntro}General remarks}
In this appendix we will refer to the different detectors as G for GEO600 \cite{GEO600}, V for VIRGO
\cite{initialVirgo,advancedVirgo}, H1 and H2 for the two LIGO detectors in Hanford (4km, 2km) and L1 for LIGO Livingston
\cite{initialLIGO,advancedLIGO}.

We will use the common abbreviations CW for continuous gravitational waves, SFT for Short Fourier Transform, PSD for
power spectral density and UL for upper limits.

The quoted sensitivity depths in tables \ref{tbl: allsky} - \ref{tbl: targeted} can correspond to different
confidence levels, as some searches use $90\%$- and others $95\%$-confidence upper limits. This applicable confidence
level is denoted by using regular versus italic font in the tables, respectively.

For searches over many frequencies, multiple targets or for upper limits reported separately for different detectors, we
us a consistent averaging procedure using the median and median absolute deviation of Eq.~\eqref{eq:15} in order to
estimate the mean and standard deviation in an outlier-robust way.

PowerFlux and loosely-coherent searches typically give separate upper limits for circular (best) polarisation and for
the worst linear
polarization, but not the more common type of population-averaged upper limits.
There has been some work estimating conversion factors for these upper limits into into polarization-averaged
sensitivity, writing $\depth\PF \sim w\worst \depth\PF\worst$ and $\depth\PF\sim w\best \depth\PF\best$.
For example \cite{wette2012estimating} obtains the conversion factors in the ranges $w\worst\sim 1.1-1.3$ and $w\best \sim
0.39-0.46$.
More recent work estimating these conversion factors on O1 data (cf.\ Fig.\cite{O1ASE@H}) for $90\%$-confidence
upper limits yields \cite{Vladimir} $w\worst 1.51 \pm 0.13$ and $w\best = 0.52 \pm 0.02$.
However, these conversion factors were obtained by treating the set of upper limits as a whole, they should not be used
to derive a proxy of population average upper limits in individual frequency bands.
Furthermore, PowerFlux strict upper limits are derived by taking the highest upper limits over regions of parameter
space. This procedure has the advantage of the upper limits retaining validity over any subset of parameter space, such
as a particular frequency and or particular sky location. However, the maximization procedure makes it difficult to
convert the data into population average upper limits which are more robust to small spikes in the data.
Given that there is currently some uncertainty on the detailed values of the conversion factors to use for different
PowerFlux searches, here we report the best/worst upper limits converted into sensitivity depths separately in
tables~\ref{tbl: allsky} and \ref{tbl: directed}.

Generally, for converting $h_0$ upper limits into depths according to Eq.~\eqref{eq:10}, we need to use an estimate for
the corresponding noise PSD $\Sn$, for which we either use a corresponding PSD over the data used in the search, where
available, or a 'generic' PSD estimate from LIGO for the given science run \cite{iLIGO_PSDs,O1_PSDs} otherwise. This adds
another level of uncertainty in the conversions, which could easily be in the range $10\%-20\%$ due to different
calibrations and different types of averaging over time.

\subsection{\label{ssec: allsky} All-sky searches, see Table~\ref{tbl: allsky}}

\subsubsection{\search{S2}{\AllSky}{Hough} \cite{S2AS}}
\label{sec:S2-AS-Hough}
The first all-sky search for CWs from isolated neutron stars, using a semi-coherent Hough transform method
applied on Short Fourier Transforms (SFTs) of the data of length $\Tseg = \SI{30}{min}$.
The search used data from the second LIGO Science Run (S2), and the number of SFTs used in the search was $687$ from L1,
$1761$ from H1 and $1384$ from H2.

The UL sensitivity depth for this search is calculated as the mean over the
three depths for H1, L1 and H2, where each depth is computed from the respective quoted best upper-limit value
$h_0^{95\%}$ and the corresponding PSD $\Sn$ in TABLE~III of \cite{S2AS}.

\subsubsection{\search{S2}{\AllSky}{\Fstat} \cite{S2ScoX1+AS}}
\label{sec:S2-AS-Fstat}
A matched-filtering search based on the coherent (single-detector) $\FF$-statistics, using $20$ SFTs from H1 and $20$
SFTs from L1 (SFT length $\TSFT=\SI{30}{min}$). The per-detector $\FF$-statistic values were combined via a coincidence
scheme, determining the most significant candidate in each $\sim\SI{1}{Hz}$ band, which was then used for measuring the
upper limits.

The sensitivity depth for this search is calculated from the given (combined multi-detector) upper limits
$h_0^{95\%}(f)$ over the search frequency range, combined with the harmonic mean over
generic H1- and L1- PSDs for the LIGO S2 data.

The estimate was calculated with the mean loudest templates of the search given in the paper as $\FF\thr = (39.5,32.2)$
for the L1 and H1 detector, respectively, and we used an average mismatch of $\SI{0.5}{\%}$ in the H1 search and
$\SI{1}{\%}$ in the L1 search, estimated from Figs.~27,28 in \cite{S2ScoX1+AS}.

\subsubsection{\search{S4}{\AllSky}{\{StackSlide,Hough,PowerFlux\}}\cite{S4AS}}
\label{sec:S4-AS-MultiPipeline}
Three semi-coherent all-sky searches using different search methods, all based on incoherently combining SFTs of length
$\Tseg=\SI{30}{min}$.
The StackSlide and the Hough search used $1004$ SFTs from H1 and $899$ from L1 and the Hough search additionally
included $1063$ SFTs from H2. The PowerFlux search used $1925$ and $1628$ SFTs from H1 and L1, respectively.

The sensitivity depths are calculated from the quoted upper limits $h_0^{95\%}(f)$ from each of the three searches over
the search frequency range, combined with the PSDs for two (H1 and L1) detectors (as a common reference) from the S4 science
run. Note that the Hough depth corresponds to the quoted multi-detector UL, while the other searches reported
only per-detector ULs.

\subsubsection{\search{S4}{\AllSky}{\Fstat\Coinc}\cite{S4E@HAS}}
\label{sec:S4-AS-FstatCoinc}
A search which used the distributed computing project Einstein@Home \cite{EinsteinAtHome} to analyse $\SI{300}{h}$ of
H1 data and $\SI{210}{h}$ of L1 data from the S4 run. The data was split into $\SI{30}{h}$ long segments coherently
analysed with the multi-detector $\FF$-statistic followed by a coincidence-step.
The measured sensitivity depth $\depth\meas^{90\%}$ is calculated by converting the quoted sensitivity factors
$R_{90\%}=\{31.8, 33.2\}$ (for frequencies below and above $\SI{300}{Hz}$, respectively) into sensitivity depths.
However, given these were computed with respect to an (arithmetic) averaged PSD estimate (given in Fig.1 in the paper),
we first converted these factors back into equivalent $h_0$ values using the mean-PSD, and then computed the Depth with
respect to the harmonic-mean (over detectors) generic noise PSD for S4.

\subsubsection{\search{earlyS5}{\AllSky}{PowerFlux} \cite{S5Early}}
\label{sec:earlyS5-AS-PowerFlux}
An all-sky search with PowerFlux over the first eight months of S5 data.
The search in total used roughly $\SI{4077}{h}$ of H1 data and $\SI{3070}{h}$ L1 data, divided into SFT segments of
$\Tseg = \SI{30}{min}$.

The sensitivity depth is calculated from the quoted per-detector upper limits $h_0^{95\%}(f)$ over the search
frequency range and the corresponding S5 noise PSDs.

\subsubsection{\search{earlyS5}{\AllSky}{\Fstat\Coinc} \cite{S5E@HEarly}}
\label{sec:earlyS5-AS-FstatCoinc}
An all-sky search run on Einstein@Home \cite{EinsteinAtHome}, using $\SI{660}{h}$ of data from H1 and
$\SI{180}{h}$ of L1 data, taken from the first 66 days of the LIGO S5 science run.
The data was divided into $28$ segments of $\Tseg=\SI{30}{h}$ duration, and each segment was searched using the
fully-coherent multi-detector $\FF$-statistic. These per-segment $\FF$-statistics were combined across segments using a
coincidence scheme.

The measured sensitivity depth $\depth\meas^{90\%}$ is calculated as the median over the converted sensitivity depths
converted from the quoted sensitivity factors $R_{90\%} = \{29.4, 30.3\}$ in the paper for the frequencies below and above $\SI{400}{Hz}$, respectively.

\subsubsection{\search{S5}{\AllSky}{PowerFlux} \cite{S5AS}}
\label{sec:S5-AS-PowerFlux}
An all-sky search using PowerFlux analyzing the whole of LIGO S5 data, broken into more than $80\,000$ $50\,\%$-overlapping
$30$-minute SFTs from both H1 and L1.

The sensitivity depth is calculated from the quoted upper limits $h_0^{95\%}$ and the S5 noise PSD.

\subsubsection{\search{S5}{\AllSky}{\HoughF} \cite{S5R5}}
\label{sec:S5-AS-HoughF}
An all-sky search using the \HoughF{} variant of the semi-coherent Hough method described in
Sec.~\ref{sec:hough-ff:-summing}, which was run on Einstein@Home.
The analyzed data consisted of $5550$ and $5010$ SFTs from the LIGO H1 and L1 interferometers, respectively, taken from
the second year of the S5 science run.
The data was divided into $121$ segments of length $\Tseg=\SI{25}{h}$, and the coherent per-segment $\FF$-statistic was
combined via the Hough method to compute the Hough number count of Eq.~\eqref{eq:24}.

The sensitivity depth of the search is calculated from the quoted $h_0^{90\,\%}$ upper limits and the corresponding S5
noise PSD.

The estimated sensitivity depth uses the generalization of the estimator described in Sec.~\ref{sec:hough-tracking-ff}
with a number-count threshold of $\ncTH=70$, a per segment threshold of $\FCO\thr = 2.6$ and a mismatch histogram
obtained from an injection-recovery simulation (with an average mismatch of $\misCO = 0.61$).

\subsubsection{\search{S5}{\AllSky}{Hough} \cite{S5Hough}}
\label{sec:S5-AS-Hough}
An SFT-based Hough all-sky search on S5 data.
The search was split into the first and the second year of S5, which were searched separately.
The first year used $11\,402$ SFTs from H1, $12\,195$ SFTs from H2 and $8\,698$ SFTs from L1, of length $\TSFT=\SI{30}{min}$.
The analysis of the second year used $12\,590$ H1-SFTs, $12\,178$ H2-SFTs and $10\,633$ L1-SFTs.

The sensitivity depth is calculated from the quoted $h_0^{90\%}$ upper limits of the second year search found in the paper and from
the S5 noise PSD.

\subsubsection{\search{S5}{\AllSky}{\GCT} \cite{S5GC1HF}}
\label{sec:S5-AS-EaH}
A high frequency all-sky search to complement previous lower-frequency all-sky searches on S5 data.
The search used the so-called GCT method \cite{2009PhRvL.103r1102P} implementing the \StackSlideF{} statistic and was
run on the distributed Einstein@Home platform.
The search used a total of $17\,797$ SFTs spanning the whole two years of S5 data from H1 and L1, divided into $205$
segments of length $\Tseg = \SI{30}{h}$.

The measured sensitivity depth $\depth\meas^{90\%}$ is determined by extrapolating the depth values given in the paper
for critical ratios of $0$ and $3.5$ to the median critical ratio over all frequency bands of $-0.15$ according to
figure 6 of \cite{S5GC1HF}.

For the estimate we determined the median threshold over all frequency bands from figure 4 of \cite{S5GC1HF} to
$2\avgFstat\thr = 5.72$. Two mismatch histograms at $\SI{1255}{Hz}$ and $\SI{1495}{Hz}$ generated with
injection-recovery studies were used. The average mismatch for both was  $\mu \approx 0.82$. The quoted value is the
mean of the two estimates with different mismatch histograms.

\subsubsection{\search{VSR1}{\AllSky}{\TDFstat} \cite{VSR1}:}
\label{sec:VSR1-AS-TDFstat}
An all-sky search using data from the first Virgo science run, VSR1. The search method uses a time-domain implementation
of the coherent $\FF$-statistic, computed over $2$-day coherent segments, which are combined using coincidences.
In total the search used $\SI{134}{days}$ of data.

The measured sensitivity depth $\depth\meas^{90\%}$ is calculated as median of the given sensitivity factors of $15.6$ and $22.4$.

\subsubsection{\search{\{VSR2,4\}}{\AllSky}{FreqHough+FUP} \cite{VSR24LF}}
\label{sec:VSR24-AS-FreqHough}
This all-sky search was performed using data from initial Virgos second (VSR2) and forth (VSR4) science run. It used the
FrequencyHough transform as incoherent step with $149$ days of data of VSR2 and $476$ days of data of VSR4 using
segments of length $\SI{8192}{seconds}$. The initial candidates were followed-up using $10$ times longer segments.

The measured sensitivity depth was calculated from upper limits $h_0^{90\%}$ extracted from figure 12 of \cite{VSR24LF}
and the harmonic mean of the PSD estimates of VSR2 and VSR4 in $\SI{0.1}{Hz}$ frequency bands.

\subsubsection{\search{S6}{\AllSky}{\GCT} \cite{S6Bucket}}
\label{sec:S6-AS-EaH}
This search used $12\,080$ SFTs from L1 and H1 data to perform a \StackSlideF{} search based on the GCT implementation,
and was run on Einstein@Home.
The search used $90$ coherent segments of length $\Tseg = \SI{60}{h}$.

The measured sensitivity depth $\depth\meas^{90\%}$ is determined by extrapolating the depth from the given critical ratios
$0$ and $6$ to the median critical ratio of $-0.07$ according to figure 5 of \cite{S6Bucket}.

The estimated depth is given for a threshold of $2\avgFstat\thr = 6.694$ which is the median of the thresholds given
for the frequency bands in figure 4 of \cite{S6Bucket}.
For the estimate two mismatch histogram created with injection-recovery studies for $\SI{55}{Hz}$ and $\SI{505}{Hz}$
was used. The average mismatch of the grid in the parameter space was at both frequencies found to be $\mu = 0.72$. The
quoted value is the mean of the two estimates with different mismatch histograms.

\subsubsection{\search{S6}{\AllSky}{\GCT+\FollowUp} \cite{S6BucketFU}}
\label{sec:S6-AS-EaH-FUP}
A multi-stage follow-up on candidates from the \search{S6}{\AllSky}{\GCT} search described in the previous paragraph,
zooming in on candidates using increasingly finer grid resolution and longer segments.
Every candidate from the initial stage with $2\avgFstat \ge 6.109$ was used as the center of a new search box for the
first-stage follow-up, continuing for a total of four semi-coherent follow-up stages.
The sensitivity of the search is dominated by the initial-stage threshold, because the later stages are designed to have a
very low probability of dismissing a real signal.
The measured sensitivity depth $\depth\meas^{90\%}=\SI{46.9}{\perRootHz}$ of this search is directly taken from the quoted
value in the paper.

The estimated multi-stage sensitivity of Sec.~\ref{sec:multi-stage-stacksl} using the thresholds given in the paper,
namely $\{2\avgFstat\thr\stage{i}\} = (6.109, 6.109, 7.38, 8.82, 15)$ and a mismatch histogram generated by recovery
injection studies for the main search and mismatch histograms provided by the original authors for every stage with
average mismatches $\{\mu\stage{i}\} = (0.72, 0.55, 0.54, 0.29, 0.14)$, yields a value of $\depth^{90\%} =
\SI{38.3}{\perRootHz}$, which differs significantly from the quoted measured sensitivity depth.
As discussed in Sec.~\ref{sec:accuracy}, we trace this discrepancy to the low threshold used, which significantly
affects the loudest-candidate mismatch approximation used in the theoretical estimate.

\subsubsection{\search{S6}{\AllSky}{PowerFlux} \cite{S6CompPF}}
\label{sec:S6-AS-PowerFlux}
The data used by this search span a time of $\SI{232.5}{d}$ with duty factor of the detectors of $53\%$ for H1 and
$51\%$ for L1.

The measured sensitivity depth is calculated from the quoted upper limits $h_0^{95\%}$ in the paper and the S6 noise
PSD.

\subsubsection{\search{O1}{\AllSky}{\GCT} \cite{O1ASE@H}}
\label{sec:O1-AS-EaH}
A low-frequency all-sky search for gravitational waves from isolated neutron stars using the distributed computing project
Einstein@Home on data from Advanced LIGO's first observing run (O1). This search used the GCT implementation of the
semi-coherent \StackSlideF{} method with $\Nseg = 12$ segments of length $\Tseg = \SI{210}{h}$ in the initial search stage.
The analyzed data consisted of $4\,744$ SFTs from the H1 and the L1 detector. The search also included a hierarchical
follow-up similar to the S6Bucket follow-up search\cite{S6BucketFU}.

The measured sensitivity depth $\depth\meas^{90\%}=\SI{48.7}{\perRootHz}$ of this search is directly taken from the quoted value in the paper.

The sensitivity estimate used a threshold $2\avgFstat\thr =14.5$ which we inferred from figure 4 in \cite{O1ASE@H}
and we obtained the mismatch histograms of the template grid at different frequencies using an injection-recovery
study, which yielded an average
mismatch of $\misSC = 0.35$ and $\misSC = 0.37$ at $\SI{20}{Hz}$ and $\SI{100}{Hz}$ respectively. The quoted depth is
the average of the two different estimates resulting for each mismatch histogram.
Note that the contrary to the measured sensitivity, the estimate only uses the first-stage parameters in this case, as
we currently cannot model the line-robust statistic used in the follow-up stages. However, as mentioned in
Sec.~\ref{sec:multi-stage-stacksl}, the overall detection probability is dominated by the first stage, while subsequent
stages mostly serve to reduce the false-alarm level.

\subsubsection{\search{O1}{\AllSky}{\{PowerFlux,Hough,\TDFstat\}} \cite{O1AS,O1ASFullBand}}
\label{sec:O1-AS-MultiPipeline}
Two papers detailing the results of all-sky searches on O1 data using four different search methods.

The first paper \cite{O1AS} searched the lower frequency range $[20,475]\,\Hz$, using four methods: PowerFlux,
FrequencyHough, SkyHough and a time-domain $\FF$-statistic search with segment-coincidences (denoted as \TDFstat{}).
The PowerFlux, FrequencyHough and SkyHough search used SFT lengths in the range $1800-7200\si{s}$ as coherent segments
while the Time-Domain $\FF$-statistic used a coherence time of $\Tseg =\SI{6}{d}$. The total amount of analyzed data was
about $\SI{77}{d}$ of H1 data and $\SI{66}{d}$ of L1 data.

In the second paper \cite{O1ASFullBand} three of these searches were extended up to $2000\,\Hz$, namely PowerFlux,
SkyHough and a time-domain $\FF$-statistic search with segment-coincidences (denoted as \TDFstat{}), using the same
data.

The sensitivity depths for the four searches are calculated from the quoted $h_0^{95\%}$ amplitude upper
limits and the noise PSD for the O1 science run.

Note that for the SkyHough method a sensitivity depth of $\SI{24.2}{\perRootHz}$ is quoted in the paper. However, this
value is based on a slightly different convention for the multi-detector noise PSD $\Sn$ (maximum over detectors
instead of the harmonic mean) than used here. For consistency with the other searches in table \ref{tbl: allsky} we
therefore compute the sensitivity depth by converting from the quoted $h_0^{95\%}$ upper limits instead.

A comparison of \emph{PowerFlux} $90\%$-confidence upper limits for an \emph{isotropic polarization} population were
provided for the O1 Einstein@Home paper \cite{O1ASE@H}, with a frequency spacing of $\SI{0.0625}{\Hz}$, which are
converted into sensitivity depth using the O1 noise PSD.

\subsection{\label{ssec: directed} Directed Searches, see Tables~\ref{tbl: directed},~\ref{tbl: nineyoung}}

\subsubsection{\search{earlyS5}{Crab}{\Fstat} \cite{S5Crab}}
\label{sec:S5-Crab-Fstat}
This search aimed at the Crab pulsar and used the first nine month of initial LIGO's fifth science run (S5).
It consisted of both a targeted (described in Sec.~\ref{sec:earlyS5-Crab-BayesPE}) and a directed
$\FF$-statistic search
described here.
The directed search used $182$, $206$ and $141$ days of data from the H1, H2 and L1 LIGO detectors, respectively.
The measured depth value is calculated from the given upper limits $h_0^{95\%}$ and the PSD estimate of the S5 data at
the search frequency.

The estimated depth uses the StackSlide estimator for a coherent search with $\Nseg = 1$ segment, a threshold of
$\Fth = 37$ and a maximal template bank mismatch of $5\%$ (given in the paper), from which we estimate the average
mismatch as $\misCO\sim\frac{1}{3}\,5\%$ (assuming a square lattice).

\subsubsection{\search{S5}{CasA}{\Fstat} \cite{S5CasA,2008CQGra..25w5011W}}
\label{sec:S5-CasA-Fstat}
The first search for continuous gravitational waves from the Cassiopeia A supernova remnant using data from initial
LIGO's fifth science run (S5).
The search coherently analyzed data in an interval of $12$ days ($934$ SFTs of length $\SI{30}{min}$) using the
$\FF$-statistic.

The measured sensitivity depth is obtained from the quoted upper limits $h_0^{95\%}$ in the paper and the S5 noise PSD.

The estimate is calculated using the StackSlide estimator for a coherent search ($\Nseg = 1$ segment), with the mismatch
histogram for an $A_n^*$ lattice with maximal mismatch of $\mis = 0.2$ (obtained from \textsc{LatticeMismatchHist()} in
\cite{octapps-JOSS}), and the average threshold of $2\Fth = 55.8$ (averaged over the respective loudest
$2\FF$-candidates found in each of the upper-limit bands).

\subsubsection{\search{S5}{GalacticCenter}{\GCT} \cite{S5GalacticCenter,behnke2015postprocessing}}
\label{sec:S5-GC-GCT}
The first search for continuous gravitational waves directed at the galactic center. The search used LIGO S5 data and
the GCT implementation of the \StackSlideF{} semi-coherent search algorithm with $630$ segments, each spanning
$\SI{11.5}{h}$, for total data set of $21\,463$ SFTs of length $\SI{30}{min}$.

The segments of the search were selected from the whole S5 science run in such a way as to maximize the SNR for
fixed-strength GW signals at the skyposition of the galactic center. Therefore the selected segments fall at times where
the antenna patterns of the LIGO detectors are better than average for this particular skyposition.
As discussed in Sec.~\ref{ssec: CompDirected}, the sensitivity-estimation method presented in this work assumes the
antenna patterns are averaged over multiple days, which causes a unusually large deviation between the estimate and the
measured sensitivity depth from the $h_0^{90\%}$ upper limits.

The estimate is calculated using the mismatch histogram (with mean $\misSC = 0.13$) obtained from an injection-recovery
study on the template bank of this search, and a detection threshold of $2\avgFstat\thr = 4.77$.

\subsubsection{\search{VSR4}{\{Vela,Crab\}}{5-vector} \cite{VSR4VelaCrab}}
\label{sec:VSR4-VelaCrab-5vector}
This coherent narrow-band search on the data from initial Virgo's forth science run (VSR4) was directed at the Vela and
the Crab pulsars. This search used the 5-vector method, and covers a range of $\pm\SI{0.02}{Hz}$ the twice the known
frequencies of Vela and Crab. The total amount of data used is $\SI{76}{d}$.

The measured sensitivity depth for this search was obtained from the published $h_0^{95\%}$ upper limits and the noise
PSD estimate for VSR4.

\subsubsection{\search{S6}{NineYoung}{\Fstat} \cite{S6NineYoung}}
\label{sec:S6-NineYoung-Fstat}
This search was directed at nine different targets, listed in Table~\ref{tbl: nineyoung}, each corresponding to a
(confirmed or suspected) compact object in a young supernova remnant.
The search uses a fully-coherent $\FF$-statistic. The amount of data used for every target varies between
$\SI{7.3e5}{s}$ and $\SI{3.1e6}{s}$ (cf. table \ref{tbl: nineyoung}).

The measured depth is calculated for each of the targets from the quoted upper limits $h_0^{95\%}$ and the corresponding PSD
for the actual data used in the search.

\newcommand{\avgTwoFthNineYoung}{\{58.0, 56.3, 55.6, 55.6, 55.3, 53.7, 52.8, 56.6, 54.1, 57.6\}}

The estimate for each target is calculated using the StackSlide estimator for a coherent search ($\Nseg = 1$ segment),
with the mismatch histogram for an $A_n^*$ lattice with maximal mismatch of $\mis = 0.2$ (obtained from
\textsc{LatticeMismatchHist()} in \cite{octapps-JOSS}), and the average $2\Fth$ threshold found for each
target (averaged over the respective loudest $2\FF$-candidates found in each of the upper-limit bands)
are given in table \ref{tbl: nineyoung}.

The 'NineYoung' entry in Table~\ref{tbl: directed} presents the median depth over all targets for the measured and
estimated depths, respectively.

\subsubsection{\search{S6}{CasA}{\GCT} \cite{S6CasA}}
\label{sec:S6-CasA-EaH}
A search directed at Cassiopeia A, which was run on the distributed computing project Einstein@Home using data from the
LIGO S6 science run. The search was based on the GCT implementation of the semi-coherent \StackSlideF{} statistic, with
$\Nseg=44$ segments of length $\Tseg = \SI{140}{h}$, and a total amount of data of $13\,143$ SFTs of length
$\SI{30}{min}$ from the two LIGO detectors in Hanford (H1) and Livingston (L1).
The measured sensitivity depth given in table~\ref{tbl: directed} is computed from the $h_0^{90\%}$ upper limits
quoted the paper \cite{S6CasA} combined with the corresponding PSD estimates.
However, as discussed in \ref{ssec: CompDirected}, this measurement suffered from a bug in the upper-limit script and
as a result is somewhat too conservative (i.e.\ too high).

The estimated sensitivity is calculated assuming an average threshold of $\avgFstat\thr = 8.25$  (estimated from Fig.~4
in \cite{S6CasA}) using the mean over estimates with different mismatch histograms generated by injection-recovery
studies at different frequencies (spanning $50-1000\,\Hz$, average mismatch $\sim 9\%$).

\subsubsection{\search{S6}{OrionSpur}{LooselyCoherent} \cite{S6OrionSpur}}
\label{sec:S6-OrionSpur-PowerFlux}
This was a search employing the so-called loosely-coherent method, aimed at the Orion spur towards both the inner and outer regions of our Galaxy.
The explored sky regions are disks with $\SI{6.87}{^\circ}$ diameter around $20^h10^m54.71^s + 33^\circ33'25.29''$ and
$\SI{7.45}{^\circ}$ diameter around $8^h35^m20.61^s-46^\circ49'25.151''$. The data used in this search spanned
$\SI{20085802}{s}$ with duty factors of $53\%$ and $51\%$ for LIGO Hanford and Livingston respectively. Due to weighting
of the data the effective amount of data used was only $\sim 12.5\%$ of the available S6 data. For the analysis data
segments of length $\SI{30}{min}$ were searched coherently.

The measured sensitivity depth was calculated from the quoted upper limits $h_0^{95\%}$ and a PSD estimate for the LIGO
S6 data.

\subsubsection{\search{S6}{NGC6544}{\Fstat} \cite{S6GlobClust}}
\label{sec:S6-NGC6544-Fstat}
This was the first search directed at the nearby globular cluster NGC~6544.
The search coherently analyzed data from the two LIGO detectors S6 science run with the $\FF$-statistic, using a single
coherent segment with $\Tseg = \SI{9.2}{d}$. The search analyzed two different data stretches separately.
The first one contained $374$ SFTs while the second contained $642$ SFTs, with SFT duration of $\SI{30}{min}$.

The measured depth was determined from the upper limits $h_0^{95\%}$ given in figure 2 of \cite{S6GlobClust} and a PSD
estimate for the LIGO S6 run.

The estimate used the StackSlide estimator with one segment, a threshold of $2\Fth = 55$ (quoted in the paper) and an
average mismatch of $0.2/3$ (assuming a roughly square lattice).

\subsubsection{\search{O1}{Narrow-band}{5-vector}\cite{O1Narrowband}}
\label{sec:O1-NarrowBand-5vector}
A narrow-band search aiming at 11 known pulsars using the fully-coherent 5-vector method on data from Advanced LIGO's
first observing run (O1).
The search used a total of $121$ days of data from the Hanford (H1) and Livingston (L1) detectors.

The sensitivity depth in the table is calculated from the median over the single-target depths, which are converted from
the upper-limits $h^{95\%}$ quoted in the paper and the corresponding noise PSD of the data used.

\subsubsection{\search{O1}{\{SN1987,GalacticCenter\}}{Radiometer}\cite{O1RadiometerDirected}}

Described in Sec.\ref{sec:O1-Radiometer}.

\subsection{\label{ssec: binary} Searches for neutron stars in binary systems, see Table~\ref{tbl: binary}}

\subsubsection{\search{S2}{ScoX1}{\Fstat}\cite{S2ScoX1+AS}}
\label{sec:S2-ScoX1-Fstat}
This first search designed specifically aimed at the NS in the LMXB system Scorpius X-1, using a coherent
single-detector \Fstat-statistic and a coincidence check on a $\SI{6}{h}$ long stretch of S2 dat.

The measured sensitivity depth was calculated from the quoted upper limits $h_0^{95\%}$ in the paper (for the
zero-eccentricity case $e=0$) and the PSD estimate of the corresponding S2 data.

\subsubsection{\search{S5}{ScoX1}{Sideband}\cite{S5ScoX1}}
\label{sec:S5-ScoX1-Sideband}
A search aimed at Scorpius X-1 by incoherently combining sidebands of a coherent \Fstat-statistic search that only
demodulates the signal for the sky-position but not its binary-orbital Doppler modulation.
This method used a stretch of $10$ days of data selected from the S5 science run for maximal sensitivity.
Two searches were performed, one with no prior assumptions about the orientation of Sco-X1, and one using more
restrictive angle-priors based on electromagnetic observations.

Bayesian upper limits $h_0^{95\%}$ were computed over the search frequency range, which we convert into sensitivity
depths (for the unknown-polarization case, see Fig.5(a) in \cite{S5ScoX1}) using the noise PSD for the data given in the
paper. In each 1Hz-band, $2\times10^6$ upper limit values were quoted, of which we use the maximum value in each
1Hz-band in order to be consistent with the usual ``loudest-candidate'' approach of setting upper limits in a given
frequency band.

\subsubsection{\search{\{S6,VSR2,3\}}{\{AllSky,ScoX1\}}{TwoSpect}\cite{S6Binary}}
\label{sec:S6VSR23-ASScoX1-TwoSpect}
A TwoSpect search for unknown binary signals from any sky-position, and a directed TwoSpect search for Scorpius X-1
specifically. This search used data from LIGO S6 science run, as well as from Virgo VSR2 and VSR3 runs, spanning
$\SI{40551300}{s}$ from each detector.

The quoted upper limits $h_0^{95\%}$ for the all-sky search and the Scorpius X-1 search were converted into Depths using
a combined (generic) PSD for the S6, VSR2 and VSR3 science runs.

\subsubsection{\search{S6}{\{ScoX1,J1751\}}{TwoSpect}\cite{S6ScoX1}}
\label{sec:S6-ScoX1-TwoSpectGrant}
A search for CW from the low-mass X-ray binaries Scorpius X-1 and XTE J1751-305 using the TwoSpect algorithm. It used
about $\SI{4e7}{s}$ from each of the two detector in the S6 science run. It used two different length of the SFTs
$\SI{840}{s}$ and $\SI{360}{s}$ which also where the length of the coherently analysed segments.

The given sensitivity depth $\depth_0^{95\%}$ is obtained from the quoted $h_0^{95\%}$ upper limits combined with the
corresponding noise PSD for S6 data.

\subsubsection{\search{O1}{ScoX1}{Viterbi}\cite{O1ScoX1}}
\label{sec:O1-ScoX1-Viterbi}
A search aimed at Scorpius X-1 using the Viterbi search method performed on $130$ days of data from Advanced LIGO's
first observational run (O1), segmented into coherent segments of length $\Tseg=\SI{10}{days}$.

The measured sensitivity depth is converted from the quoted upper limits $h_0^{95\%}$ (for unknown polarization) and the
noise PSD of the corresponding O1 data.

Note that contrary to many other search methods, this search setup appears to result in a frequency-dependent
sensitivity depth, namely $\depth(f)\propto f^{-1/4}$ (see Eq.(9) in \cite{O1ScoX1}). For consistency with other searches, we quote the median and
(MAD) standard-deviation over frequencies in Table~\ref{tbl: binary}, and note that the total range of sensitivity depths
of this search is found as $\depth(f)\sim 11\,(f/f_0)^{-1/4}\,\si{\perRootHz}\in[ 4.6, 11.2 ]\,\si{\perRootHz}$ 
with $f_0=60.5\,\Hz$.

\subsubsection{\search{O1}{ScoX1}{CrossCorr}\cite{O1ScoX1CrossCorr}}
\label{sec:O1-ScoX1-CrossCorr}
This search aimed at Scorpius X-1 using the CrossCorr search algorithm using data from Advanced LIGO's first
observational run (O1). The data was split into coherently analysed segments (SFTs) with a (frequency-dependent) length
between $\SI{240}{s}$ and $\SI{1400}{s}$.

The measured sensitivity depth is obtained from the quoted (isotropic-prior) upper limits $h_0^{95\%}$  and the noise
PSD of the O1 data. Note, however, that the search ULs are given per $0.05\,\Hz$ bands, which is unusually small
compared to most other upper-limit bands (typically $0.25-1\,\Hz$), and therefore they display more variability.
In order to make these ULs more comparable to other searches, we use the 95th-percentile highest upper limits per
1Hz-bands (as recommended in Fig.~5 of \cite{O1ScoX1CrossCorr}).
This 'binning' procedure only has a small effect on the resulting sensitivity depth, which is reduced from
$\SI{25.3}{\perRootHz}$ to $\SI{24.0}{\perRootHz}$.

Note that this search has a frequency-dependent sensitivity depth, which starts at around
$\depth(25\,\Hz)\sim \SI{45}{\perRootHz}$ for low frequencies, asymptoting down to $\depth\sim \SI{23}{\perRootHz}$
above $f\gtrsim 800\,\Hz$.
However, in order to be consistent with other searches, we quote the median and (MAD) standard deviation over all
frequencies in Table.~\ref{tbl: binary}.

\subsubsection{\search{O1}{\{ScoX1 and others\}}{Radiometer}\cite{O1RadiometerDirected}}
\label{sec:O1-Radiometer}

The 'Radiometer' search method, which was developed mainly for stochastic background searches, can also be used for
directed CW searches at particular sky-positions. This method does not use a particular signal model, which allows it to
be sensitive to a wide range of possible signal families, at the cost of somewhat lower sensitivity to 'regular' CW
signals.
This search aimed at the sky-positions of Sco-X1, as well as at the supernova remnant 1987A and the galactic center.

The search reported $h_0^{90\%}$ (and $h_0^{95\%}$ for Sco-X1, reported in \cite{O1ScoX1CrossCorr}) upper limits in
narrow frequency bands of $1/32\Hz = 0.03125\,\Hz$ bands, which is unusually small compared to most other upper-limit bands
(typically $0.25-1\,\Hz$), and therefore they display more variability. In order to make these ULs more comparable to
other searches, we use the 95th-percentile highest upper limits per 1Hz-bands (as recommended in Fig.~5 of
\cite{O1ScoX1CrossCorr}), and following the same procedure as used for the CrossCorr results (discussed in
Sec.~\ref{sec:O1-ScoX1-CrossCorr}).

\subsection{\label{ssec: targeted} Targeted Searches, see Table~\ref{tbl: targeted}}

\subsubsection{\search{S1}{J1939+21}{\{\Fstat,\BayesPE\}}\cite{S1J1939}}
\label{sec:S1-J1939-FstatBayesPE}
This first CW search on data from GEO\,600 and LIGO's first science run (S1).
It used $(16.7, 5.73,8.73,8.9)$ days of data from four detectors, GEO\,600 (G1), LIGO Livingston (L1), LIGO
Hanford-\SI{4}{km} (H1), and LIGO Hanford-\SI{2}{km} (H2), respectively.
Two types of searches were performed, a coherent \Fstat-statistic search as well as direct Bayesian parameter
estimation (\BayesPE).

Table \ref{tbl: targeted} gives the mean and standard deviation for the sensitivity depths over the four detectors.
The measured sensitivity depth for the \Fstat-search was determined from the quoted upper limits $h_0^{95\%}$ in
table~IV\cite{S1J1939} for the most pessimistic $\iota$ ($\cos\iota=0$) and $\psi$, and from the quoted numbers in the conclusion for
the (standard) population-averaged orientation. The noise PSD values are taken from table~III in \cite{S1J1939}.
The corresponding estimate is calculated with the StackSlide estimator for $\Nseg = 1$ and quoted threshold values
$2\Fth = (1.5,3.6,6.0,3.4)$ for the four detectors from table~III in the paper.
For the 'worst-case' estimate we use the prior $\cos \iota = 0$ and minimise the sensitivity depth over
$\psi\in[-\pi/4,\pi/4]$ in order to reflect the 'conservative' ULs quoted in the paper.
Note, however, that contrary to the typically small false-alarm level (p-value) of the UL thresholds used (typically
$1\%$), the loudest candidates used here as thresholds here had relatively high p-values of $83\%$, $46\%$, $20\%$ and
$49\%$, respectively, as seen in table~III of \cite{S1J1939}.

\subsubsection{\search{S2}{Known pulsars}{\BayesPE}\cite{S2KnownPulsar}}
\label{sec:S2-KnownPulsars-BayesPE}
A coherent targeted search for $28$ known isolated radio pulsars was performed using the Bayesian parameter-estimation
pipline (\BayesPE) on data from the second LIGO Science Run (S2), using $\SI{910}{h}$ of data from H1, $\SI{691}{h}$
from H2 and $\SI{342}{h}$ of L1 data from the S2 data set.

The measured sensitivity depth is calculated from the quoted Bayesian upper limits $h_0^{95\%}$ and corresponding noise
PSD estimates for the S2 science run.

The sensitivity estimate is performed using the Bayesian sensitivity estimator, for simplicity assuming the sources are
distributed isotropically over the sky.

\subsubsection{\search{\{S3,4\}}{Known pulsars}{\BayesPE}\cite{S3S4KnownPulsar}}
\label{sec:S34-KnownPulsars-BayesPE}
This search targeted $78$ known radio pulsars by analysing $(45.5, 42.1, 13.4)$ days of data from the three
detectors (H1, H2, L1) from the third science run (S3) of LIGO and GEO\,600, and $(19.4, 22.5, 17.1)$ days of data from
the three detectors from the S4 science run. The analysis used the Bayesian parameter-estimation pipeline (\BayesPE).

The measured sensitivity depth was determined from the quoted Bayesian upper limits $h_0^{95\%}$ combined with the noise
PSD of the S3 and S4 science runs combined (using harmonic mean).

The sensitivity estimate is calculated using the Bayesian sensitivity estimate, for simplicity assuming the sources to be
isotropically distributed on the sky.

\subsubsection{\search{earlyS5}{Crab}{\BayesPE}\cite{S5Crab}}
\label{sec:earlyS5-Crab-BayesPE}

This search on 9 months of data from the early LIGO S5 science run targeted only the Crab pulsar at twice its rotation
rate, using the Bayesian parameter-estimation pipeline. A corresponding narrow-band search using the \Fstat-statistic is
described in Sec.~\ref{sec:S5-Crab-Fstat}.
The targeted search used $201$, $222$ and $158$ days of data of the H1, H2 and L1 LIGO detectors.

The measured depth is determined from the quoted (i.e.\ the corrected value in the Erratum) upper limit $h_0^{95\%}$
assuming an isotropic polarization prior, and the corresponding noise PSD of the detectors for the early S5 science run data.

\subsubsection{\search{S5}{Known pulsars}{\BayesPE}\cite{S5KnownPulsar}}
\label{sec:S5-KnownPulsars-BayesPE}
A search targeting 116 known pulsars using $525$ days of H1 data, $532$ days of H2 data and $437$ days of L1 data
from LIGO's fifth science run (S5). The search employed the Bayesian parameter-estimation pipeline.

The measured sensitivity depth is calculated from the quoted Bayesian upper limits $h_0^{95\%}$ and the noise PSD of the
S5 data.

The estimate is calculated with the Bayesian sensitivity estimator under the assumption that the targets are distributed
isotropically over the sky.

\subsubsection{\search{VSR2}{Vela}{\{\BayesPE,\Fstat,5-vector\}}\cite{VSR2Vela}}
\label{sec:VSR2-Vela-MultiMethod}
A targeted search for the Vela pulsar using Virgo's second science-run (VSR2) data, using three different methods:
Bayesian parameter estimation, the \Fstat-statistic (and \Gstat-statistic) and the 5-vector method. The data set
consisted of $149$ days of Virgo data.

Two types of searches and upper limits were computed, namely (i) using uninformative (isotropic) priors on the pulsar
orientation, and (ii) using angle priors on $\cos\iota$ and $\psi$ from electromagnetic observations.

In table \ref{tbl: targeted} we only give the measured depth corresponding to the isotropic prior, averaged over the
three methods, which obtained very similar results. This was computed from the quoted upper limits $h_0^{95\%}$ and the
noise PSD for the Vela VSR2 run.
The measured sensitivity depth obtained when using the angle priors is found as $\depthVelaAnglePriors\,\si{\perRootHz}$.

The estimated sensitivity depth is calculated using the Bayesian sensitivity estimator.

\subsubsection{\search{\{S6,VSR2,4\}}{Known pulsars}{\{\BayesPE,\Fstat,5-vector\}}\cite{S6KnownPulsar}}
\label{sec:S6VSR24-KnownPulsar-MultiMethod}
This search targeted 195 known pulsars, using $149$ days of VSR2 and $76$ days of VSR4 data for pulsars with a
CW frequency lower than $f< \SI{40}{Hz}$ and an additional $238$ days of S6 data from H1 and $225$ days from L1 for faster
spinning pulsars with $f>\SI{40}{Hz}$. The analysis was done using three different methods: Bayesian parameter
estimation, the \Fstat-statistic (or \Gstat-statistic for restricted angle priors) and the 5-vector method.

The given measured sensitivity depth in table~\ref{tbl: targeted} is the median and MAD standard deviation over the
sensitivity depths for the different targets (averaged over high- and low-frequency targets). The sensitivity depths are
obtained from the quoted upper limits $h_0^{95\%}$ and the corresponding noise PSD estimate of the data used (which is
either S6 and VSR2 and VSR4 for high-frequency targets $f>\SI{40}{Hz}$, or only VSR2 and VSR4 for low-frequency
targets).

The estimated sensitivity is obtained from the Bayesian sensitivity estimator assuming an isotropic prior over the sky,
averaged over high- and low-frequency depths results.

\subsubsection{\search{O1}{Known pulsars}{\{\BayesPE,\Fstat,5-vector\}}\cite{O1KnownPulsar}}
\label{sec:O1-KnownPulsar-MultiMethod}
In this search $200$ known pulsars were targeted using three different methods: Bayesian parameter estimation, the
\Fstat-statistic (or \Gstat-statistic for restricted angle priors) and the 5-vector method.
The searches used $78$ and $66$ days of H1 and O1 data from the first observational run of advanced LIGO (O1),
respectively.

The measured sensitivity depth is obtained from the quoted Bayesian upper limits $h_0^{95\%}$ over all targets and the
corresponding noise PSD for the LIGO detectors during O1.

The estimated sensitivity depth is determined from the Bayesian estimator as an all-sky estimate assuming the targets
are isotropically uniformly distributed over the sky.

\section{CW Signal model and $\FF$-statistic}
\label{sec:cw-signal-model}

A plane gravitational wave arriving from a direction $\vSky$ (unit vector) can be written
\cite{1996BonazzolaGourgoulhon} in TT gauge (in the notation of \cite{2014CQGra..31f5002W}) as a purely spatial strain
tensor $\tensor{h}$ with two polarizations $+,\times$, namely
\begin{equation}
  \label{eq:38}
  \tensor{h}(\tau) = h\plus(\tau)\,\tensor{e}\plus + h\cross(\tau)\,\tensor{e}\cross\,,
\end{equation}
where $\tau$ is the emission time of the signal in the source frame, and $\tensor{e}\plus$ and
$\tensor{e}\cross$ are the two polarization basis tensors, which can be constructed from a right-handed orthonormal
basis $\{\uvec{\ell}, \uvec{m}, -\vSky\}$ as $\tensor{e}\plus = \uvec{\ell}\otimes\uvec{\ell} - \uvec{m}\otimes\uvec{m}$ and
$\tensor{e}\cross = \uvec{\ell}\otimes\uvec{m} +  \uvec{m}\otimes\uvec{\ell}$.


The measured scalar CW signal $h^X(t)$ at time $t$ by detector $X$ is the response of the detector to the GW tensor
$\tensor{h}(\tau^X(t))$, where $\tau^X(t)$ denotes the emission time of a wavefront that reaches detector $X$ at time $t$.
This timing relationship depends on the sky-position $\vSky$ of the source as well as any binary-orbital parameters in
case of a CW from a neutron star in a binary system, as it describes the time-dependent light-travel time from the
source to the detector.
In the long-wavelength limit we assume the GW wavelength to be much larger than the detector armlength, which is a good
approximation for current ground-based detectors up to $\si{\kilo\hertz}$ frequencies.
This allows us to write the detector response as a tensor contraction (in both tensor indices):
\begin{equation}
  \label{eq:43}
  h^X(t) = \tensor{d}^X(t) : \tensor{h}(\tau^X(t))\,,
\end{equation}
where $\tensor{d}^X = \uvec{u}\otimes\uvec{u} - \uvec{v}\otimes\uvec{v}$ for interferometer arms along unit vectors
$\uvec{u}$ and $\uvec{v}$.

It is helpful to define a \emph{source-independent} orthonormal polarization basis
$\{\uvec{\imath},\uvec{\jmath},-\vSky\}$ instead, where for any sky position $\vSky$, the unit vector $\uvec{\imath}$ is
chosen to lie in Earth's equatorial plane (pointing West) and $\uvec{\jmath}$ is pointing in the northern hemisphere.
This defines the (sky-position dependent) alternative polarization basis as
$\tensor{\eps}\plus(\vSky)\equiv \uvec{\imath}\otimes\uvec{\imath} - \uvec{\jmath}\otimes\uvec{\jmath}$ and
$\tensor{\eps}\cross(\vSky)\equiv \uvec{\imath}\otimes\uvec{\jmath} + \uvec{\jmath}\otimes\uvec{\imath}$.
The rotation between these two basis systems defines the \emph{polarization angle} $\psi$, which is measured
counterclockwise from $\uvec{\imath}$ to $\uvec{\ell}$, and relates the two polarization basis tensors as
\begin{align}
  \label{eq:42}
  \tensor{e}\plus  &= \hspace{0.7em} \tensor{\eps}\plus\,\cos2\psi + \tensor{\eps}\cross\,\sin2\psi\,\\
  \tensor{e}\cross &= -\tensor{\eps}\plus\,\sin2\psi + \tensor{\eps}\cross\,\cos2\psi\,.
\end{align}
Combining these expression, we can obtain the factored signal form
$h^X(t; \Amp,\lambda) = \Amp^\mu\,h^X_\mu(t; \lambda)$ of Eq.~\eqref{eq:36}, which
was first derived in \cite{JKS1998data}.
The four amplitudes $\{\Amp^\mu\}_{\mu=1}^4$ depend on the signal amplitude $h_0$, the inclination angle
$\iota$, polarization angle $\psi$, and the reference-time phase $\phi_0$, namely
\begin{equation}
  \label{eq:37}
  \begin{split}
    \Amp^1 &= \hspace{0.7em}A\plus\,\cos\phi_0\,\cos2\psi - A\cross\,\sin\phi_0\,\sin2\psi\,,\\
    \Amp^2 &= \hspace{0.7em}A\plus\,\cos\phi_0\,\sin2\psi + A\cross\,\sin\phi_0\,\cos2\psi\,,\\
    \Amp^3 &= -  A\plus\,\sin\phi_0\,\cos2\psi - A\cross\,\cos\phi_0\,\sin2\psi\,,\\
    \Amp^4 &= -  A\plus\,\sin\phi_0\,\sin2\psi + A\cross\,\cos\phi_0\,\cos2\psi\,,
  \end{split}
\end{equation}
and the four (detector-dependent) basis functions $h_\mu^X(t;\lambda)$ are
\begin{equation}
  \label{eq:44}
  \begin{split}
    h_1^X(t) &= a^X(t)\, \cos\phi(\tau^X(t))\,,\\
    h_2^X(t) &= b^X(t)\, \cos\phi(\tau^X(t))\,,\\
    h_3^X(t) &= a^X(t)\, \sin\phi(\tau^X(t))\,,\\
    h_4^X(t) &= b^X(t)\, \sin\phi(\tau^X(t))\,,
  \end{split}
\end{equation}
in terms of the \emph{antenna-pattern functions} $a^X(t),b^X(t)$ given by the contractions
\begin{equation}
  \label{eq:45}
  \begin{split}
    a^X(t; \vSky) &= \tensor{d}^X(t) : \tensor{\eps}\plus(\vSky)\,,\\
    b^X(t; \vSky) &= \tensor{d}^X(t) : \tensor{\eps}\cross(\vSky)\,.
  \end{split}
\end{equation}
Using the factored signal form of Eq.~\eqref{eq:36}, the log-likelihood ratio Eq.~\eqref{eq:53} now takes the form
\begin{equation}
  \label{eq:53}
  \ln \Lambda(x; \Amp, \lambda) = \Amp^\mu\,x_\mu - \frac{1}{2}\Amp^\mu \M_{\mu\nu} \Amp^\nu\,,
\end{equation}
where we defined
\begin{equation}
  \label{eq:50}
  x_\mu(\lambda) \equiv \scalar{x}{h_\mu},\quad\text{and}\quad
  \M_{\mu\nu}(\lambda)  \equiv \scalar{h_\mu}{h_\nu}\,,
\end{equation}
in terms of the four basis function $h_\mu(t;\lambda)$ defined in Eq.~\eqref{eq:44}.
The $4\times4$ \emph{antenna-pattern matrix} $\M$ can be shown to be well approximated by the block-diagonal form
\begin{equation}
  \label{eq:51}
  \M = \Sn^{-1}\,\Tdata\,
  \begin{pmatrix}
    M & 0\\
    0 & M\\
  \end{pmatrix}
  \;\text{with}\;
  M \equiv \begin{pmatrix}
    A & C\\
    C & B\\
  \end{pmatrix}\,,
\end{equation}
defining the antenna-pattern coefficients $A,B,C$, which depend on the sky-position $\vSky$.

We see in Eq.~\eqref{eq:53} that the log-likelihood ratio is a quadratic function of the amplitudes
$\Amp^\mu$, and can therefore be analytically maximized \cite{JKS1998data} (or marginalized \cite{2009CQGra..26t4013P})
to yield the well-known $\FF$-statistic:
\begin{equation}
  \label{eq: Fstat}
  \begin{split}
    \FF(x;\lambda) &\equiv \max_{\Amp} \ln \Lambda(x;\Amp, \lambda)\\
                   & = \frac{1}{2}\, x_\mu\,\M^{\mu\nu}\,x_\nu\,,
  \end{split}
\end{equation}
with $\M^{\mu\nu}$ defined as the inverse matrix to $\M_{\mu\nu}$ of Eq.~\eqref{eq:51}.

\section{Distribution of $\FF$-statistic maximized over correlated templates}
\label{sec:distr-ff-stat}

It has been a long-standing assumption (e.g.\ \cite{S5CasA,wette2012estimating} that the distribution of the statistic
$2\Fmax(x)\equiv\max_{\lambda_i}2\FF(x;\lambda_i)$ in Gaussian noise $x$, maximized over a template bank
$\lambda_i\in\templBank$ of $i = 1\ldots\Ntempl$ (generally correlated) templates can be modelled by assuming
maximization over an ``effective'' number of uncorrelated trials $\Ntempl'$ instead, namely
\begin{equation}
  \label{eq:56}
  \prob{2\Fmax}{\Ntempl'} = \Ntempl'\,\cdf_0(2\Fmax)^{\Ntempl'-1}\,\pdf_0(2\Fmax)\,,
\end{equation}
where
\begin{align}
  \label{eq:57}
  \pdf_0(2\FF) &= \prob{2\FF}{\rho=0}\,,\\
  \cdf_0(2\FF) &= \int_0^{2\FF} \pdf_0(2\FF')\,\DD2\FF'\,,
\end{align}
where the (single-template) $\FF$-statistic in pure Gaussian noise follows a central $\chi^2$ distribution (with four
degrees of freedom in the fully-coherent case Eq.~\eqref{eq:13}, or $4\Nseg$ degrees of freedom for a semi-coherent
$\FF$-statistic over $\Nseg$ segments, Eq.~\eqref{eq:12}).

We show here by counter-example that the model of Eq.~\eqref{eq:56} is not generally accurate, as correlations between
templates do not simply modify $\Ntempl'$ but also change the functional \emph{form} of the distribution.
It has been hypothesized previously \cite{wette2012estimating} that these (already-observed) deviations might be due to
certain approximations (c.f.\ \cite{prix2010f}) used in the numerical implementation of the $\FF$ statistic. While such
effects will account for some amount of deviation, one can show this effect to be quite small overall.

We demonstrate the fundamental statistical nature of this discrepancy by using a simpler example:
we generate a time-series $\{x_j\}_{j=0}^{N-1}$ of $N = 200$ samples drawn from a Gaussian distribution and compute the
Fourier transform $\tilde{x}_k$ normalized to $E[|\tilde{x}_k|^2] = 2$, such that $2\FF_2(x,f) \equiv |\tilde{x}(f)|^2$
follows a central $\chi^2$ distribution with two degrees of freedom in every frequency bin $f$. We can therefore set
$\pdf_0(2\FF_2)=\chi^2_2(2\FF_2;0)$ and use the corresponding cdf in Eq.~\eqref{eq:56}.

\begin{figure*}[htbp]
  \begin{tabular}{l l}
    (a) & (b)\\
    \includegraphics[width=0.5\textwidth]{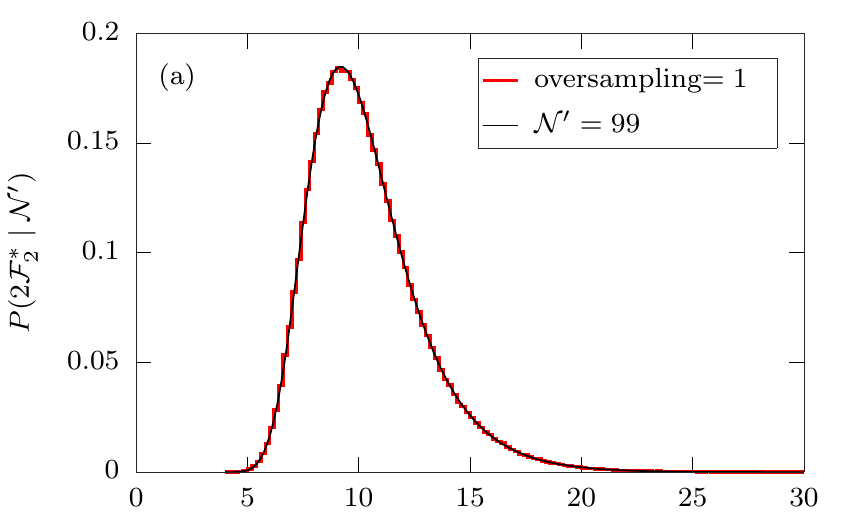}
        & \includegraphics[width=0.5\textwidth]{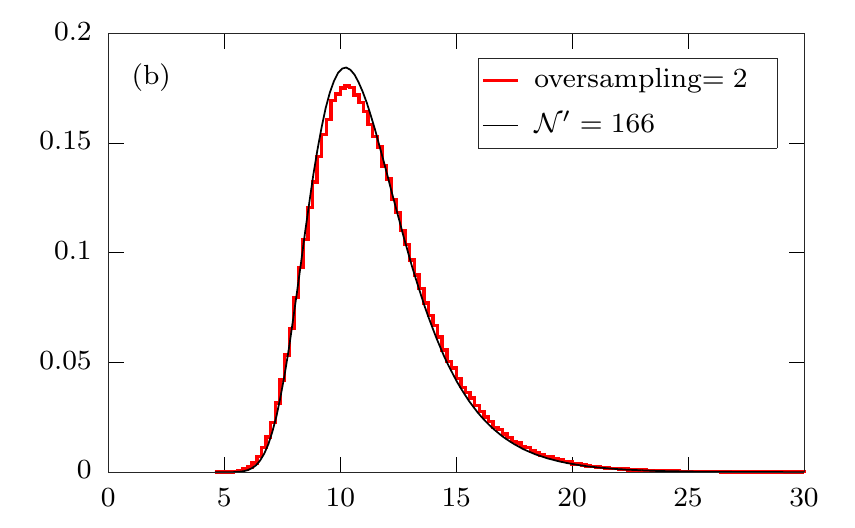} \\
    (c) & (d)\\
    \includegraphics[width=0.5\textwidth]{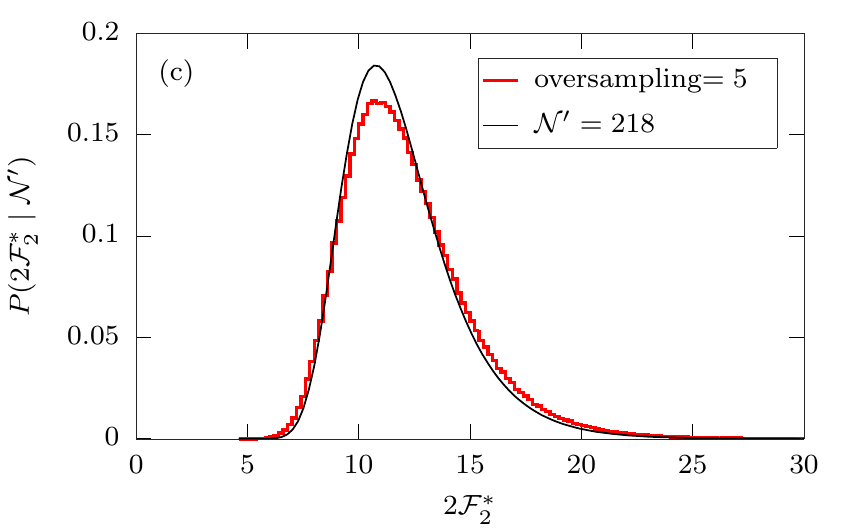}
    & \includegraphics[width=0.5\textwidth]{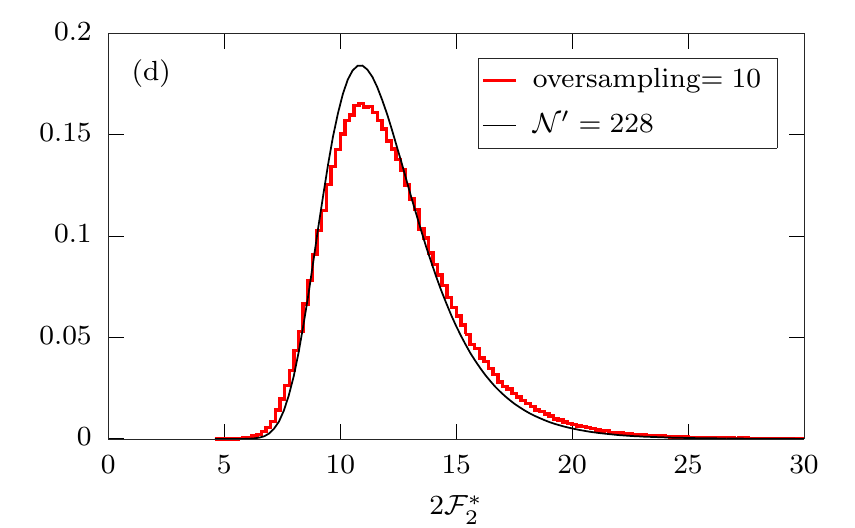}\\
  \end{tabular}
  \caption{\emph{Stair-case plot}: histogram (over $10^6$ repeated trials) of $2\Fmax_2=\max_{k}|\tilde{x}_k|^2$ for Fourier
    transforms of Gaussian-noise timeseries, using different oversampling factors (a)--(d), where $\text{oversampling}=1$
    corresponds to the original FFT frequency resolution.
    \emph{Solid thin line}: corresponding best-fit theoretical model Eq.~\eqref{eq:56} with an effective number of
    templates $\Ntempl'$.}
  \label{fig:FFT-max-pdf}
\end{figure*}

We consider different cases of \emph{oversampling} by zero-padding the time-series to a multiple (denoted
as the oversampling factor in Fig.~\ref{fig:FFT-max-pdf}) of the original $N$ time samples:
the $N/2-1=99$ (positive) frequency bins without oversampling are strictly uncorrelated (and we also know that there
can be at most $N=200$ independent templates in total, given the length of the initial timeseries). With increasing
oversampling, the correlations between frequency bins increase. We repeate this process $10^6$ times for different noise
realizations, and in each case we compute $2\Fmax_2(x)$ over all the (positive) frequency bins of the Fourier power, and
histogram these values. We then fit the number of effective templates $\Ntempl'$ in the theoretical distribution of
Eq.~\eqref{eq:56} by minimizing the (symmetric) Jensen–Shannon divergence between the measured and theoretical
distributions. The results are shown in Fig.~\ref{fig:FFT-max-pdf} for different cases of oversampling.
We see that for increased oversampling, i.e.\ more correlations between 'templates' (i.e.\ frequency bins), the
functional form of the histogram agrees less with the theoretical distribution assuming independent templates. The
effect seems to saturate for oversampling $\gtrsim 10$, with $\Ntempl\sim230$ greater than the known maximal number (i.e.\
$N=200$ of (strictly) independent template in this vector space.

There is no simple or intuitive explanation for this effect that we are aware of, but it is reminiscent of a similarly
surprising result found in the localization of the maximum over different assumed signal durations of transient CW
signals, see Figs.~8 and 9 in \cite{prixetal2011:_transientCW}. The distribution of the statistic is identical in each
time-step, but the steps are correlated, resulting in a peculiar non-uniform distribution of the location of the maximum.

\end{appendix}

\bibliography{paper}

\end{document}